\documentclass[journal, 10pt]{IEEEtran}
\IEEEoverridecommandlockouts
\usepackage[utf8]{inputenc}
\usepackage{amssymb,amsfonts}
\usepackage{amsthm}
\usepackage{graphicx}
\usepackage{epstopdf}
\usepackage{textcomp}
\usepackage{xcolor}
\usepackage{bbm}
\usepackage[english]{babel}
\usepackage{caption}
\usepackage{subcaption}
\usepackage{floatrow}
\usepackage{algorithm} 
\usepackage{algorithmic}
\usepackage{amsmath}
\usepackage{multicol}
\usepackage{multirow}
\usepackage{gensymb}
\usepackage{float}
\usepackage{tikz}
\usepackage{pgfplots}
\usepackage{array,ragged2e}
\usepackage{latexsym} 
\usetikzlibrary{patterns}
\usetikzlibrary{shapes, arrows, arrows.meta}
\usepackage{setspace}
\newtheorem{theorem}{Theorem}
\newtheorem{definition}{Definition}
\newtheorem{lemma}{Lemma}

\newtheorem{problem}{Problem}
\floatstyle{plaintop}
\restylefloat{table}
\newcolumntype{P}[1]{>{\RaggedRight\arraybackslash}p{#1}}
\algsetup{linenosize=\tiny}
\DeclareMathOperator*{\argmax}{\arg\!\max}
\DeclareMathOperator*{\argmin}{\arg\!\min}
\newcommand{\bb}{{\bf {b}}}

\newcommand{\bu}{{\bf {u}}}
\newcommand{\bw}{{\bf {w}}}
\newcommand{\bG}{{\bf {G}}}

\newcommand{\mcA}{{\mathcal{A}}}
\newcommand{\mcB}{{\mathcal{B}}}
\newcommand{\mcE}{{\mathcal{E}}}
\newcommand{\mcF}{{\mathcal{F}}}
\newcommand{\mcL}{\mathcal{L}}
\newcommand{\mcM}{\mathcal{M}}
\newcommand{\mcN}{\mathcal{N}}
\newcommand{\mcP}{\mathcal{P}}
\newcommand{\mcR}{\mathcal{R}}
\newcommand{\mcS}{\mathcal{S}}
\newcommand{\mcU}{\mathcal{U}}
\newcommand{\mcV}{\mathcal{V}}
\newcommand{\mcX}{\mathcal{X}}
\newcommand{\mcZ}{\mathcal{Z}}
\newcommand{\mbB}{\mathbb{B}}
\newcommand{\mbE}{\mathbb{E}}
\newcommand{\mbL}{\mathbb{L}}
\newcommand{\mbM}{\mathbb{M}}
\newcommand{\mbN}{\mathbb{N}}
\newcommand{\mbU}{\mathbb{U}}
\newcommand{\mbV}{\mathbb{V}}
\newcommand{\mbW}{\mathbb{W}}
\newcommand{\mbX}{\mathbb{X}}
\newcommand{\mbY}{\mathbb{Y}}

\newcommand\indicator{{\mathbbm{1}}}
\newcommand\topt{\mathrm{opt}}
\newcommand\tde{\mathrm{de}}
\newcommand\tdn{\mathrm{dn}}
\newcommand\tth{\mathrm{th}}
\newcommand\tut{\mathrm{ut}}
\newcommand\ag{\mathrm{ag}}
\newcommand\tac{\mathrm{ac}}
\newcommand\tsr{\mathrm{sr}}

\newcommand\ta{\mathrm{a}}
\newcommand\tb{\mathrm{b}}
\newcommand\tc{\mathrm{c}}
\newcommand\td{\mathrm{d}}

\newcommand\tg{\mathrm{g}}

\newcommand\tn{\mathrm{n}}
\newcommand\tp{\mathrm{p}}
\newcommand\ts{\mathrm{s}}
\newcommand\tu{\mathrm{u}}
\newcommand\algDLIG{Algorithm 3}

\pgfplotsset{width=7cm,compat=newest}
\pgfplotsset{
label style={font=\small},
legend style={font=\footnotesize},
}
\pgfplotsset{compat=1.13,
    /pgfplots/ybar legend/.style={
    /pgfplots/legend image code/.code={%
       \draw[##1,/tikz/.cd,yshift=-0.25em]
        (0cm,0cm) rectangle (5pt,1.0em);},
   },
   }

\makeatletter
\def\endthebibliography{%
	\def\@noitemerr{\@latex@warning{Empty `thebibliography' environment}}%
	\endlist
}
\makeatother
\begin{document}
	\author{Santosh Kumar Singh, Satyabrata Sahu, Ayushi Thawait, Prasanna  Chaporkar, and Gaurav S. Kasbekar}
        \title{Joint User and Beam Selection in Millimeter Wave Networks 
	%
        }
	\maketitle
{\renewcommand{\thefootnote}{} \footnotetext{S.K. Singh,  P. Chaporkar, and G.S. Kasbekar are with the Department of Electrical Engineering, Indian Institute of Technology (IIT) Bombay, Mumbai, India. S. Sahu is with  Radio Firmware, Connectivity Solution, Samsung Semiconductor, India. A. Thawait is with the Airports Authority of India (AAI), Ministry of Civil Aviation, Government of India. Their email addresses are santoshiitb@ee.iitb.ac.in,  chaporkar@ee.iitb.ac.in, gskasbekar@ee.iitb.ac.in, satyabrata.s@samsung.com, and ayushi.t@aai.aero, respectively. SS and AT worked on this research while they were at IIT Bombay. The work of SKS, PC, and GSK is supported in part by the project with code RD/0121-MEITY01-001. 
	
\par  A part of this paper was presented at the Fourteenth International Conference on Ubiquitous and Future Networks (ICUFN) 2023 and was published in its proceedings (DOI: 10.1109/ICUFN57995.2023.10199880)~\cite{singh2023ngub}. 
\par This work has been submitted to the IEEE for possible publication. Copyright may be transferred without notice, after which this version may no longer be accessible.
}}
\begin{abstract}
 We study the problem of selecting a user equipment (UE) and a beam for each access point (AP) for concurrent transmissions in a millimeter wave (mmWave) network, such that the sum of weighted rates of UEs is maximized. We prove that this problem is NP-complete. We propose two algorithms-- Markov Chain Monte Carlo (MCMC) based and local interaction game (LIG) based UE and beam selection-- and prove that both of them asymptotically achieve the optimal solution. Also, we propose two fast greedy algorithms-- NGUB1 and NGUB2-- for UE and beam selection. Through extensive simulations, we show that our proposed greedy algorithms outperform the most relevant algorithms proposed in prior work and perform close to the asymptotically optimal algorithms. 
\end{abstract}
	

\section{Introduction} \label{Section: Introduction}
 The volume of data traffic exchanged using wireless networks is increasing rapidly, due to an increase in demand for voice and data services~\cite{cisco2020}, which the existing sub-6 GHz cellular bands are unable to meet even using advanced techniques such as massive Multiple Input Multiple Output (MIMO) and heterogeneous networking~\cite{andrews2014what,agiwal2016next}. To meet this demand, millimeter wave (mmWave) bands can be used, since they have a large amount of unutilized spectrum available and also have the potential to provide multi-gigabit data rates~\cite{niu2015survey}. But because of the high carrier frequency, mmWave communications suffer from significant propagation loss. Therefore, beamforming needs to be used, which makes mmWave communications inherently directional. To achieve effective communication, the characteristics of mmWave networks, viz., their directional nature, short transmission range, and dense deployment of access points (APs), which give rise to various challenges including blockage, interference, and frequent handovers, need to be handled by developing novel strategies~\cite{niu2015survey,sakaguchi2017where,liu2016user,attiah2020survey}.    

To overcome the above mentioned challenges and to provide robust connectivity to user equipment (UE), several algorithms have been proposed to select beams and UEs for each AP in mmWave networks~\cite{sur2016beamspy,zhou2017beamforecast,haider2018listeer,wei2017pose,jog2019many,yang2020mmmuxing,yang2021mdsr,zhang2022reinforcement,sha2023versatile}. In~\cite{sur2016beamspy,zhou2017beamforecast,haider2018listeer}, mmWave networks with a single AP and a single UE are considered and algorithms for beam selection at the AP for the UE are designed such that the connection remains robust to blockage and UE mobility. In~\cite{wei2017pose}, mmWave networks with multiple cooperating APs and a single UE are considered and an algorithm is designed to select a beam for each AP to make a robust connection with the mobile UE. In~\cite{jog2019many,yang2020mmmuxing,yang2021mdsr}, mmWave networks with multiple APs and multiple UEs are considered and  algorithms are designed to select a beam and a UE for each AP using the measured received signal strength (RSS)  information at each UE corresponding to each beam of every AP in the network. The work in~\cite{wei2017pose,jog2019many,yang2020mmmuxing,yang2021mdsr} performs beam and  UE selection for each AP in two steps. First, they select mutually exclusive subsets of UEs for different APs. Second, they jointly select a UE and a beam for each AP, such that the UE for each AP is selected from the subset of UEs obtained for the AP in the first step.  However, this two-step selection process may not provide a weighted sum rate maximizing set of UE and beam pairs for different APs.
Also, none of the previous works analyzed the computational complexity of the problem or tried to quantify the optimal value for the problem.

In this paper, we consider mmWave networks containing multiple APs and multiple UEs.  Each AP has a set of beams, and can, at a time, communicate with at most one UE using a beam from the set. Each AP is connected to a central controller using a high-speed link. Our aim is to design an algorithm for the central controller that uses RSS information (RSSs at different UEs from different APs using different beams) and the weights of UEs, and provides a set of AP, UE, and beam triplets for concurrent data transmission such that the weighted sum rate of UEs is maximized. 

Our contributions are as follows: 
\begin{itemize}
    \item We prove that the problem of finding the optimal UE and beam for each AP in a mmWave network with multiple APs, multiple UEs, and multiple beams at each AP using RSS information to maximize the weighted sum rate of UEs is NP-complete~\cite{kleinberg2006algorithm}.
    \item We propose two algorithms-- a Markov Chain Monte Carlo (MCMC) based algorithm and a local interaction game (LIG) based algorithm--  for the UE and beam selection problem and prove that both of them asymptotically obtain the optimal solution. These two algorithms are offline algorithms and can be used to benchmark the performance of online algorithms. Also, these algorithms can be used to generate a data set consisting of a mapping of RSS information to optimal UE and beam pairs and the generated data set can be used to train online learning algorithms for UE and beam selection in mmWave networks.
    \item Through simulations, we show that the most relevant algorithms for UE and beam selection in prior works~\cite{wei2017pose,jog2019many,yang2020mmmuxing,yang2021mdsr} underperform the proposed benchmark algorithms by a large margin.
    This shows the need for designing novel algorithms.
    \item We design two fast greedy approach-based algorithms, NGUB1 and NGUB2 (Novel Greedy UE and Beam Selection), for UE and beam selection at each AP.   
    \item Through extensive simulations, we show that the weighted sum rate achieved by our proposed benchmark algorithms is the same as that achieved using exhaustive search for a small mmWave network. Also, we show that the weighted sum rate achieved by our proposed NGUB1 and NGUB2 algorithms is close to that achieved by our proposed asymptotically optimal algorithms. Further, we show that our proposed algorithms, NGUB1 and NGUB2, outperform the most relevant algorithms proposed in prior works~\cite{wei2017pose,jog2019many,yang2021mdsr} by a large margin. 
\end{itemize}

The rest of this paper is organized as follows. Section \ref{Section: Related Work} provides a review of related prior literature. Section \ref{Section: System Model and Problem Formulation} presents the system model and problem formulation. Section \ref{Section: Complexity} analyzes the computational complexity of the formulated problem. Section \ref{Section: MCMC based UE and Beam Selection} and Section \ref{Section: LIG based UE and Beam Selection} present the proposed MCMC-based and LIG-based  asymptotically optimal algorithms, respectively. Section \ref{Section: Requirement for design of new algorithm} motivates the need for the design of a new algorithm for UE and beam selection. Section \ref{Section: Proposed Algorithm} describes our proposed novel algorithms. Section \ref{Section: Performance Evaluation} presents simulation results. Finally,  Section \ref{Section: Conclusions} provides conclusions and directions for future research.

\section{Related Work} \label{Section: Related Work}

In this section, we present a brief review of some relevant prior works on resource allocation, including UE and beam selection, in sub-6 GHz and mmWave networks.
In~\cite{kim2011distributed}, a distributed $\alpha$ optimal user association algorithm for a wireless cellular network is proposed; different association rules-- load-based, delay-based, Signal to Noise Ratio (SNR)-based, and throughput-based user association-- correspond to different values of the parameter $\alpha$. 
In~\cite{madan2010cell}, a user association and resource partitioning algorithm for heterogeneous Long Term Evolution (LTE)-Advanced cellular networks is proposed for inter-cell interference coordination to improve the coverage gain and capacity. In~\cite{awais2017efficient}, an interference-aware greedy algorithm is proposed for user association and resource allocation in a cloud radio access network (CRAN) after formulating joint association and resource allocation as a combinatorial optimization problem. In~\cite{ye2013user}, several approximation algorithms for user association are proposed after modeling user association considering load balancing in HetNets as an NP-hard utility-maximizing problem. 
In~\cite{kim2011distributed,madan2010cell,awais2017efficient,ye2013user}, user association and/ or resource allocation is considered with a different objective, and for different types of sub-6 GHz networks; however, unlike our paper, none of them addresses joint user and beam selection in mmWave networks.

In~\cite{mesodiakaki2016energy}, with an objective to maximize energy and spectrum efficiency, user association in mmWave backhaul small-cell networks is formulated as non-convex power minimization with constraints on spectrum resources. The optimal Pareto frontier solution of the problem is derived and a low complexity user association scheme is proposed that attains high energy efficiency given spectral efficiency. In~\cite{khawam2020coordinated}, joint spectrum allocation and user association in 5G heterogeneous networks (HetNets) using mmWave bands is formulated as a non-cooperative potential game. A projected sub-gradient-based algorithm is proposed to solve a convex optimization problem that corresponds to finding the equilibrium point of the game. In~\cite{skouroumounis2017low}, a two-stage base station (BS) selection scheme, in which, in the first stage, a user defines a set of candidate BSs using some pre-selection scheme and in the second stage, the user associates with the candidate BS that maximizes the signal to interference and noise ratio (SINR),  for heterogeneous mmWave networks is proposed and an asymptotic expression for coverage probability is derived for each pre-selection scheme. In~\cite{zhang2017energy}, with the objective of maximizing the long-run throughput of users, a joint user association and power allocation problem in an ultra-dense mmWave network is formulated as a mixed integer programming problem and iterative gradient based user association and power allocation are proposed. In~\cite{athanasiou2013auction}, with the objective of maximizing users' throughput, user association in a 60 GHz mmWave network is formulated as a linear optimization problem, and an auction algorithm-based solution approach is proposed by transforming the formulated problem into a minimum cost flow problem. In~\cite{liu2019joint}, joint user association and resource allocation considering users' priority in mmWave HetNets is formulated as a mixed-integer non-linear programming (MINLP) problem, and using Lagrangian dual decomposition, an iterative algorithm is proposed to obtain the  user association and power allocation. In~\cite{soleimani2018cluster}, with the objective of maximizing the sum-rate of a cluster, clustering-based resource allocation and user association in mmWave femtocell networks is formulated as a binary optimization clustering problem and solved by converting it into a continuous optimization problem. 
In~\cite{mesodiakaki2016energy,khawam2020coordinated,skouroumounis2017low,zhang2017energy,athanasiou2013auction, liu2019joint,soleimani2018cluster}, user association and/ or resource allocation are considered with a different objective, and for different types of  mmWave networks; however, unlike our paper, none of them considers joint user and beam selection in mmWave networks with multiple APs and multiple UEs.

In~\cite{sur2016beamspy}, an algorithm called Beamspy is proposed to select an alternative beam for a UE whenever the current beam quality degrades without costly beam searching. It exploits channel sparsity and spatial correlation information available at the AP to predict the outage of the current beam and to suggest a new beam. In~\cite{zhou2017beamforecast}, an algorithm called Beamforecast is proposed, which predicts the beam at the AP for a mobile UE in real time without costly beam scanning.  It exploits spatial correlation in channel profiles for prediction. In~\cite{haider2018listeer}, an algorithm called Listeer is proposed, which steers the beam at the AP towards a mobile UE. It acquires direction estimates for beams at APs using indicator Light Emitting Diodes (LEDs) on APs and off-the-shelf light sensors at UEs. In~\cite{wei2017pose}, beam selection at multiple cooperating APs for a mobile UE using pose information-- location and orientation of the UE-- is proposed. The scheme maps pose information of the mobile UE with measured link quality and selects a beam based on pose information. In contrast to the work in~\cite{sur2016beamspy,zhou2017beamforecast,haider2018listeer,wei2017pose}, where beam selection is done for a network consisting of a  single UE, our work considers the problem of UE and beam selection in a mmWave network having multiple APs and multiple UEs.

In~\cite{xu2019ping}, a ping-pong-like optimization method using hybrid particle swarm optimization and simulated annealing is proposed to obtain a set of UEs and beams for selected UEs for concurrent transmission in a downlink mmWave network consisting of an AP and multiple UEs. In~\cite{hegde2019matching}, a matching theory-based algorithm that outputs a beam for each UE and corresponds to the stable matching of the formulated matching problem is proposed in a downlink mmWave network consisting of an AP having multiple radio frequency (RF) chains and multiple UEs with the number of RF chains being greater than the number of UEs. In~\cite{cheng2020low}, the problem of UE and beam pair selection in a downlink mmWave network consisting of an AP having multiple RF chains and multiple UEs is modeled as a stable matching problem and a low computational complexity algorithm is proposed to select a set of UE-beam pairs for concurrent transmission. In~\cite{ahn2022machine}, a machine learning-based low overhead beam selection scheme for UEs is proposed in a downlink indoor mmWave network consisting of an AP equipped with a single camera and multiple RF chains, and multiple UEs. The proposed algorithm takes an image from the camera as input, estimates the angle of each UE, and provides a beam such that the sum rate of concurrent transmissions can be maximized. However, unlike our paper, none of~\cite{xu2019ping,hegde2019matching,cheng2020low,ahn2022machine} considers UE and beam selection in an mmWave network with multiple APs and multiple UEs.

In~\cite{alizadeh2022reinforcement,zhang2021non,wang2023joint,jog2019many,yang2020mmmuxing,yang2021mdsr}, the considered mmWave networks have multiple APs and multiple UEs. In~\cite{alizadeh2022reinforcement}, reinforcement learning-based user association and handover are proposed for two-tier HetNets consisting of both sub-6 GHz and mmWave APs. The proposed algorithm aims to balance the load across APs. In~\cite{zhang2021non},  a non-cooperative game-based distributed beam scheduling scheme for coordinated multipoint (CoMP) networks, consisting of Wi-Fi and mmWave APs operating in unlicensed spectrum, is proposed.  In~\cite{wang2023joint}, an algorithm that maximizes the number of users successfully served in both access and backhaul links,  for joint user association and transmission scheduling in integrated mmWave access and terahertz backhaul networks is proposed. The algorithm associates a UE to a BS based on the ratio of the rate obtained by the UE and its minimum rate requirement and schedules transmissions based on the minimum slot requirement by associated UEs. 
 In~\cite{jog2019many}, a scheme called  BounceNet is proposed, which selects a beam and UE for each AP by mapping the problem to a conflict graph and the selection of beam and UE corresponds to the selection of a weighted maximum independent set. In~\cite{yang2020mmmuxing}, a scheme called mmMuxing is proposed for user and beam selection using channel measurements in each schedule. It can determine the user and beam for each AP that leads to the minimum interference in each schedule. In~\cite{yang2021mdsr}, a scheme called MDSR is proposed for user and beam selection and it is an extension of the work in~\cite{yang2020mmmuxing}. The authors of~\cite{yang2021mdsr} also quantify the impact of interference on network performance through measurement and conclude that the prediction-based interference minimizing approach~\cite{sur2016beamspy,zhou2017beamforecast,wei2017pose} in existing work is inefficient. However, none of the above works analyzes the complexity of the problem of UE and beam selection  to maximize the sum rate of the UEs in an mmWave network having multiple UEs and multiple APs. Also, none of the above works provides an asymptotically optimal algorithm to solve this problem. In contrast, in this paper, we consider mmWave networks containing multiple APs and multiple UEs. We prove that the problem of joint UE and beam selection is NP-complete~\cite{kleinberg2006algorithm}. We propose two asymptotically optimal algorithms to solve this problem. We also propose two novel algorithms that are robust to UE mobility and the presence of blockages in the network. Using extensive simulations, we show that they are close to optimal and that they outperform the algorithms proposed in prior work~\cite{wei2017pose,jog2019many,yang2021mdsr} by a large margin.

\section{System Model and  Problem Formulation}
\label{Section: System Model and Problem Formulation}
\subsection{System Model }
\begin{figure}[!ht]
    \centering
    \includegraphics[scale=0.2]{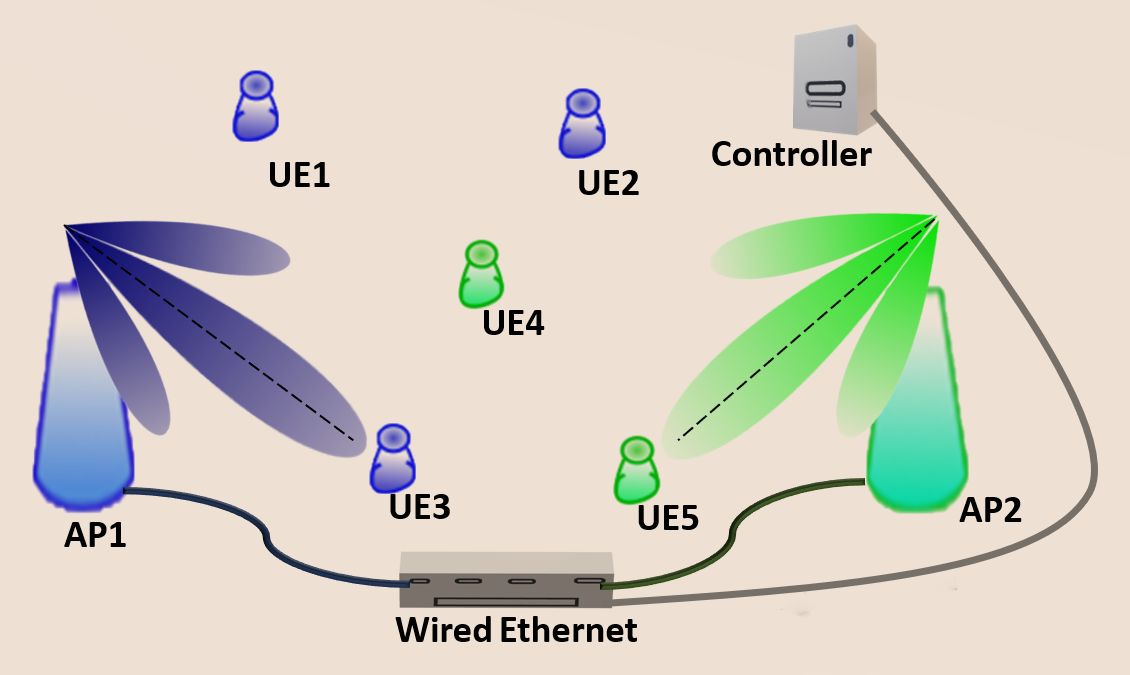}
    \caption{This is an illustration of our system model for an example scenario in which there are two APs and five UEs. $\operatorname{UE_{3}}$ and $\operatorname{UE_{5}}$ are concurrent users.}
    \label{Fig:System Model}
\end{figure}
We consider a downlink mmWave network with $N_A$ APs and $N_U$ UEs. Each AP has $N_B$ beams, and it can communicate with one UE at any instant of time using one of its beams. Also, at any instant of time, a UE can communicate with at most one AP. Each AP in the network is linked to a central controller via a high-capacity wired Ethernet backhaul link. Time is divided into slots of equal duration; additionally, a slot
$t \in \{0, 1, 2, \cdots \}$ is considered to be the duration of time from  time instant $tT$ to time instant $(t+1)T$.  Each slot is further divided into $K$ schedules of equal duration. We assume that a UE can move only at the beginning of a slot and remains static during the slot.  So the RSS at each UE from each AP and for each of its beams is computed only at the start of each slot. The RSS information computed at the beginning of a slot is used for UE and beam selection in each of the $K$ schedules of the slot. Our goal is to determine a UE and a beam for each AP for concurrent transmission in each schedule.  In this work, we have used the statistical channel model proposed for mmWave networks in~\cite{samimi20163d} to obtain RSS information. Fig. \ref{Fig:System Model} illustrates our system model.

\subsection{Problem Formulation}
Let $\mcB:=\{1,\cdots, N_B\}$, $\mcU:=\{1,\cdots, N_U\}$, and $\mcA:=\{1,\cdots, N_A\}$ denote the set of beams at each AP, set of UEs, and set of APs, respectively. Let ${\tb}_a \in \mcB$ (respectively, ${\tu}_a \in \mcU$) denote the beam (respectively, UE) chosen for AP $a \in \mcA$ in a schedule. Let $S_{b,u,a}$ denote the RSS obtained at UE $u$ when beam $b$ is used at AP $a$. Let $\mcS:=\{S_{b,u,a}|b \in \mcB,u \in \mcU, a \in \mcA\}$ denote the RSS information of a slot. Note that this information remains the same for each schedule of the slot. Let $w_u$ denote the weight of a UE in a schedule. The weight of a UE can be assigned in various ways, e.g., based on how urgently data needs to be transmitted to the UE, equal to the quantity of data that is waiting in a queue for transmission to the UE, or selected so as to ensure fairness across different UEs. Note that the weight of a UE can be changed in each schedule of a slot to ensure fairness across UEs. Let $\bw$, whose $u$th component is $w_u$, denote the weight vector of UEs in a schedule.  Let $\mcF_a=\{({\tb}_a,{\tu}_a)\}$ be a set that contains the beam and UE pair that is assigned to an AP $a$ in a schedule. Note that $\mcF_a$ can be the null set in two ways. First, $N_A >N_U$ and all UEs $u \in \mcU$ are scheduled to be served by APs other than $a$. Second, $S_{b,u,a}<S_t, \forall b \in \mcB, u \in \mcU$, where $S_t$ denotes an RSS threshold below which communication is not possible. Let $\Breve{\mcA}:=\{a \in \mcA| \mcF_a\neq \emptyset\}$ (respectively, $\mcF:=\{ ({\tb}_a,{\tu}_a,a)|a \in \Breve{\mcA}\}$) denote the set of APs (respectively, beam-UE-AP triplets) selected for transmission in a schedule. The sum of weighted rates of concurrent transmissions in a schedule given $\mcS$, $\bw$, and $\mcF$  of the schedule is: $R_{\mcS,\bw,\mcF}=$
\begin{align} 
  &\sum_{a \in \Breve{\mcA}} w_{{\tu}_a} B \log_2 \Bigg( 1+ \frac{S_{{\tb}_a,{\tu}_a,a}}{N_0+\sum_{a'\in \Breve{\mcA}, a' \neq a}S_{b_{a'},{\tu}_a,a'}} \Bigg). \label{Eq: Objective Main}
\end{align}
In the above equation, $N_0$, $B$, and $w_{{\tu}_a}$ represent noise power, bandwidth, and the weight of UE ${\tu}_a$, respectively.  Also, note that the cardinality of $\Breve{\mcA}$ can be less than $N_A$. The rate of each AP-UE-beam triplet $(\tb_a,\tu_a,a)$ in (\ref{Eq: Objective Main}) is assumed to be the Shannon capacity of the channel corresponding to the AP, UE, and beam in the triplet.  Our objective is to find a set of AP-UE-beam triplets for concurrent transmission that maximizes the sum of weighted rates.  
\begin{problem} \label{Problem: Original UE and beam selection Problem}    Find a set of AP-UE-beam triplets, $\mcF$, for a schedule that maximizes the sum of weighted rates in (\ref{Eq: Objective Main}) given $\mcS$ and $\bw$ of the schedule.
\end{problem}

\section{Complexity} \label{Section: Complexity}
\begin{theorem} \label{Theorem: The problem is NP-hard}
    Problem \ref{Problem: Original UE and beam selection Problem} is NP-complete.
\end{theorem}

We outline here the proof of Theorem \ref{Theorem: The problem is NP-hard} and provide a detailed proof in the appendix. We prove that  Problem \ref{Problem: Original UE and beam selection Problem} is NP-complete by (i) showing that it lies in class NP~\cite{kleinberg2006algorithm}, and (ii) proving that the problem of finding a maximum independent set in a graph in which each node has at most three incident edges, which has been shown to be NP-complete in~\cite{fleischner2010maximum}, can be reduced to  Problem \ref{Problem: Original UE and beam selection Problem} in polynomial time.

\section{MCMC-Based UE and Beam Selection} \label{Section: MCMC based UE and Beam Selection}
In this section, we provide an MCMC-based algorithm for UE and beam selection in an mmWave network. Also, we show that the proposed algorithm asymptotically achieves the optimal solution of Problem \ref{Problem: Original UE and beam selection Problem}.

Let $\bb$ (respectively, $\bu$) be a vector of beams (respectively, UEs) whose $a$th component is the beam (respectively, UE) selected  for the $a$th AP. Let $\mbB:=\mcB^{N_A}$ (respectively, $\mbU:=\{\bu| \bu_a \in \mcU, \forall a \in \mcA; \bu_a \neq \bu_{a'} \mbox{ for } a \neq a'  \}$) be the set of all possible beam vectors (respectively, set of all possible UE vectors). Let $R_{\mcS,\bw,\bb,\bu}=$
\begin{align}
  &\sum_{a \in \mcA} w_{\bu_a} B \log_2 \Bigg( 1+ \frac{S_{\bb_a,\bu_a,a}}{N_0+\sum_{a'\in \mcA, a' \neq a}S_{\bb_{a'},\bu_a,a'}} \Bigg) \label{Eq: Objective MCMC}
\end{align}
denote the sum of the weighted rates of UEs in a schedule given $\mcS$, $\bw$, $\bb$, and $\bu$. Note that for every schedule, an algorithm for the problem takes $\mcS$ and $\bw$ of the schedule as input and provides $(\bb,\bu)$ as output. 

\subsection{Proposed MCMC-based UE and Beam Selection Algorithm}
A pseudocode of the proposed MCMC-based UE and beam selection algorithm has been provided in Algorithm \ref{Algorithm: MCMC}. Let $R_{\bb}=\max_{\bu \in \mbU} R_{\mcS,\bw,\bb,\bu}$ denote the maximum value of the sum of the weighted rates of UEs over all possible UE vectors given $\bb$, $\mcS$, and $\bw$. 
\begin{algorithm}[!ht]
    \algsetup{linenosize=\normalsize}
	\caption{MCMC-based UE and Beam Selection} 
	\begin{algorithmic}[1] \label{Algorithm: MCMC}
	\STATE \textbf{Input:} $\mcS, \bw$ // RSS information and UEs weight vector
	
	\STATE \textbf{Output:}  $(\bb^*, \bu^*)$ // Optimal beam and UE vector 
	\STATE Initialization: $i=1$ and $\bb^0$ a randomly chosen beam vector from the set $\mbB$
 	\WHILE{$i \le I_0$} 
	\STATE Find $\Tilde{\bb}$ by replacing the beam at a randomly chosen position in $\bb^{i-1}$ with a chosen random beam from $\mcB$ 
        \STATE Compute $R_{\Tilde{\bb}}$ and $R_{\bb^{i-1}}$
        \IF{ $R_{\Tilde{\bb}} > R_{\bb^{i-1}}$ } 
        \STATE Let $\bb^i=\Tilde{\bb}$
        \ELSE 
        \STATE Compute $p=e^{\alpha (R_{\Tilde{\bb}}-R_{\bb^{i-1}})}$
        \STATE Let $\bb^i=\Tilde{\bb}$ with probability $p$ and $\bb^i=\bb^{i-1}$ with probability $1-p$.
        \ENDIF
        \STATE $i=i+1$
	\ENDWHILE
        \STATE Output $\bb^*=\bb^{I_0}$ and $\bu^*=\argmax_{\bu \in \mbU} R_{\mcS,\bw,\bb^*,\bu}$
	\end{algorithmic} 
\end{algorithm}
The proposed Algorithm \ref{Algorithm: MCMC} starts with a beam vector chosen uniformly at random from $\mbB$. It updates the beam vector iteratively. Let $\bb^i, i \in \mbN$ denote the beam vector for the $i$th iteration.  In an iteration $i$, it performs the following operations in sequence to obtain an updated beam vector $\bb^i$ given $\bb^{i-1}$. It finds a new beam vector $\Tilde{\bb}$ by replacing the beam at a randomly chosen position, which corresponds to an AP, in beam vector $\bb^{i-1}$ with a beam chosen uniformly at random from $\mcB$. The algorithm computes $R_{\Tilde{\bb}}$ and $R_{\bb^{i-1}}$, the maximum values of the weighted sum rates corresponding to the new beam vector $\Tilde\bb$ and the current beam vector $\bb^{i-1}$, respectively. If the maximum value of the weighted sum rate corresponding to the new beam vector is greater than the maximum value of the weighted sum rate corresponding to the current beam vector, then the algorithm sets the current beam vector for the $i$th iteration to be equal to the new beam vector. Otherwise, it computes $p=e^{\alpha (R_{\Tilde{\bb}}-R_{\bb^{i-1}})}$, where the parameter $\alpha$ is a positive real number. It is easy to see that $p \in [0,1]$. The algorithm sets the current beam vector for the $i$th iteration to be equal to the new beam vector (respectively, the current beam vector of the $(i-1)$th iteration) with probability $p$ (respectively, $1-p$). The algorithm updates beam vectors some finite number of times, $I_0$, and then outputs the beam vector $\bb^{I_0}$ as the optimal beam vector $\bb^*$. Also, it computes the optimal UE vector $\bu^*=\argmax_{\bu \in \mbU} R_{\mcS,\bw,\bb^*,\bu}$  and output it along with $\bb^*$.

\subsection{Optimality of MCMC-based UE and Beam Selection}
In this subsection, we show that the proposed Algorithm \ref{Algorithm: MCMC} achieves the optimal UE and beam vectors asymptotically when the parameter $\alpha$ is increased slowly from $0$ to $\infty$. Note that $\argmax_{(\bb,\bu) \in \{ \mbB \times \mbU \}} R_{\mcS,\bw,\bb,\bu}=\argmax_{\bb \in \mbB}\Big(\argmax_{\bu \in \mbU} R_{\mcS,\bw,\bb,\bu} \Big)$, i.e., the maximization of weighted sum rate can be done in two steps-- first, maximization over all possible UE vectors for each beam and then over all possible beam vectors. 

We now show that given $\mcS,\bw$, and $\bb$, the set $\argmax_{\bu \in \mbU} R_{\mcS,\bw,\bb,\bu}$ can be obtained in polynomial time. Recall that a maximum weight matching\footnote{Let $G=(\mcV,\mbE)$ be an undirected graph with set of vertices $\mcV$ and set of edges $\mbE$. Let $\mbX, \mbY \subseteq \mcV$ be disjoint sets of vertices and $\mbW$ be the set of weights of the edges in $\mbE$. Then $\bG=(\mbX,\mbY,\mbE, \mbW)$ is called a weighted bipartite graph~\cite{west2001introduction} if every edge in $\mbE$ connects a vertex of $\mbX$ to a vertex of $\mbY$. 
      $\mbM \subseteq \mbE$ is called a matching in the graph $\bG$ if no two edges in $\mbM$ have a common vertex in either $\mbX$ or $\mbY$. A matching $\mbM \subseteq \mbE$  is called a maximum weight matching~\cite{west2001introduction} if the sum of the weights of the edges belonging to it is the maximum over all matchings of $\bG$.} in a weighted bipartite graph can be obtained in polynomial time~\cite{kleinberg2006algorithm}. Thus, to show that the set $\argmax_{\bu \in \mbU} R_{\mcS,\bw,\bb,\bu}$ can be found in polynomial time, it suffices to show that our problem of finding the set $\argmax_{\bu \in \mbU} R_{\mcS,\bw,\bb,\bu}$ given $\mcS, \bw$ and $\bb$ can be modeled as finding a maximum weight matching of a weighted bipartite graph. We create a weighted bipartite graph as follows. Let $\mbX$ and $\mbY$ be the sets of APs and UEs, respectively. We assume that each element of  $\mbX$ is connected with each element of $\mbY$ and that these connections represent the edges of the bipartite graph. Also, no two vertices of either $\mbX$ or $\mbY$ are connected by an edge. We choose the weight of the edge from $x \in \mbX$ to $y \in \mbY$ to be $w_{y} B \log_2 \Bigg( 1+ \frac{S_{\bb_x,y,x}}{N_0+\sum_{x'\in \mcX, x' \neq x}S_{\bb_{x'},y,x'}} \Bigg)$. Note that the weight of the edge from $x$ to $y$ represents the rate of the UE $y$ when it is associated with AP $x$ and the beam vector $\bb$ is used at APs. It is easy to see that the maximum weight matching of this bipartite graph gives a set of edges that correspond to the associations of users with APs that maximize the sum of the weighted rates of UEs when the beam vector $\bb$ is used at APs. The vector of UEs corresponding to the edges in the maximum weight matching is $\argmax_{\bu \in \mbU} R_{\mcS,\bw,\bb,\bu}$.

The previous paragraph shows that for a fixed beam vector $\bb$, the optimal UE vector can be found in polynomial time. Thus, to find the pair of optimal UE and beam vectors, we only need to design an algorithm that effectively searches over the space of beam vectors. Note that Algorithm \ref{Algorithm: MCMC} only searches over the space of beam vectors rather than the space of both UE and beam vectors, and thus has a significantly reduced search space compared to a naive MCMC-based algorithm that simultaneously searches over the space of UE and beam vectors. 
\begin{lemma} \label{Lemma: MCMC unique stationary distribution}
    Given $\alpha$, Algorithm \ref{Algorithm: MCMC} achieves the following unique stationary distribution:
    \begin{align*}
        \pi (\bb)=\frac{e^{\alpha R_{\bb}}}{\sum_{\bb' \in \mbB} e^{\alpha R_{\bb'}}}, \quad \forall \bb \in \mbB.
    \end{align*}
\end{lemma}

\begin{lemma} \label{Lemma: MCMC iterates belongs to set of optimal beam vectors}
    Let $\mbB^{\topt} \subseteq \mbB$ be the set of optimal beam vectors. Then,  $\lim_{\alpha \uparrow \infty } \pi (\bb)=\begin{cases}
        \frac{1}{|\mbB^{\topt}|}, & \forall \bb \in \mbB^{\topt},\\
        0, & \text{otherwise}.
    \end{cases}$
\end{lemma}

\begin{theorem} \label{Theorem: MCMC-based Algo asymptotically achieve optimum UE and beam selection}
    Algorithm \ref{Algorithm: MCMC} asymptotically achieves the optimal UE and beam for each AP in Problem \ref{Problem: Original UE and beam selection Problem}.
\end{theorem}

 Here, we outline the proof of Theorem \ref{Theorem: MCMC-based Algo asymptotically achieve optimum UE and beam selection} and provide a detailed proof in the Appendix. First, in Lemma \ref{Lemma: MCMC unique stationary distribution}, we show that the proposed MCMC-based UE and beam selection algorithm achieves a unique stationary distribution over the space of beam vectors. Then in Lemma \ref{Lemma: MCMC iterates belongs to set of optimal beam vectors}, we show that for a large value of the algorithm parameter $\alpha$, the above stationary distribution assigns a total probability mass of $1$ to the set of optimal beam vectors, and hence that 
 the proposed MCMC-based UE and beam selection algorithm converges to one of the optimal beam vectors with probability $1$.

\section{LIG-based UE and Beam Selection} \label{Section: LIG based UE and Beam Selection}
In this section, we model the problem of UE and beam selection in mmWave networks as an LIG and provide an LIG-based algorithm to solve it. Also, we show that the proposed algorithm asymptotically achieves the optimal solution of Problem \ref{Problem: Original UE and beam selection Problem}.
\subsection{Game Formulation}
The LIG formulation for a problem requires the identification of the players for the problem and defining the strategies, utilities, neighbors, and payoffs of the players.
Before we provide the game formulation, we describe some notations. Let ${\tb}_a^u$ denote the beam at AP $a$ for UE $u$. Let $\Tilde{b}_a^u:=\argmax_{b \in \mcB}S_{b,u,a}$ denote the best beam in terms of RSS at AP $a$ for UE $u$. In our game, an AP-UE pair acts as a player if the RSS at the UE from the AP using the best beam at the AP for the UE is greater than some threshold value. Let $\mcL:=\{1, 2, \cdots, L\}$ denote the set of players for our game;  $\mcL$ uniquely maps to the set $\{(a,u), a \in \mcA, u \in \mcU| S_{\Tilde{b}_a^u,u, a} \ge S_t\}$, where $S_t$ is the threshold value\footnote{A low (respectively, high) value of $S_t$ leads to the selection of a large (respectively, small) number of AP-UE pairs as players and hence leads to a slow (respectively, fast) convergence rate. However, $S_t$ needs to be sufficiently small to guarantee convergence of the LIG-based algorithm to the optimal solution of Problem \ref{Problem: Original UE and beam selection Problem}.}.  Note that $L \le N_A N_U$. Let $\mcZ_l=\{0,1, \cdots, N_B\}$ (respectively, $z_l \in \mcZ_l$) denote the strategy space (respectively, a strategy)  of player $l$. The strategy $0$ (respectively, $i \in \{1, 2, \cdots, N_B\}$) of a player $l$ implies that the AP-UE pair corresponding to player $l$ is not selected (respectively, selected and beam $i$ will be used) for data transmission. In this section, if we say that a player $l$ is active/ on  (respectively, off), we mean $z_l>0$ (respectively, $z_l=0$). For $\Bar{\mcL} \subseteq \mcL$, let $\mcZ_{\Bar{\mcL}}:=\{ \times_{l \in \Bar{\mcL}} \mcZ_l \}$ (respectively, ${z}_{\Bar{\mcL}} \in \mcZ_{\Bar{\mcL}}$) denote the strategy space (respectively, strategy profile)  of the players belonging to set $\Bar{\mcL}$. Let $-l:=\mcL\setminus l$ denote the set of all players except player $l$. Thus,  $\mcZ_{-l}$ (respectively, $z_{-l} \in \mcZ_{-l}$) denotes the strategy space (respectively, strategy profile) of all other players except $l$.
Let $l_{\tu}$ (respectively, $l_{\ta}$) denote the UE (respectively, AP) corresponding to player $l$. Let $l_{\tb}$ denote the beam at AP $l_{\ta}$ for UE $l_{\tu}$. Let $\mcN_l^{\ta}:=\{\Tilde{l} \in \mcL|l_{\ta}=\Tilde{l}_{\ta} \}$ (respectively, $\mcN_l^{\tu}:=\{\Tilde{l} \in \mcL|l_{\tu}=\Tilde{l}_{\tu} \}$) denote the set of players that have the same AP (respectively, UE) as player $l$ has. Let $I_l^{l'}:=S_{\Tilde{b}_{l_{\ta}}^{l_{\tu}},{l'}_{\tu},l_{\ta}}$ denote the interference created by the best beam at AP $l_{\ta}$ for UE $l_{\tu}$ to UE ${l'}_{\tu}$. Let $\mcN_l^{\tg}:=\{ l' \in \mcL|I_{l'}^{l} >I_{\tth}, l' \notin \{ \mcN_l^{\ta} \bigcup \mcN_l^{\tu} \} \}$ (respectively, $\mcN_l^{\tc}:=\{ l' \in \mcL|I_{l}^{l'} >I_{\tth}, l' \notin \{ \mcN_l^{\ta} \bigcup \mcN_l^{\tu} \} \}$) denote the set of links by (respectively, on)  which link $l$ gets interfered (respectively, creates interference) with the interference greater than some threshold value $I_{\tth}$\footnote{A low (respectively, high) value of $I_{\tth}$ leads to the selection of a large (respectively, small) number of neighbors to which a player creates interference and from which a player gets interfered and hence leads to slow (respectively, fast) convergence. However, $I_{\tth}$ needs to be sufficiently small to guarantee convergence of the LIG-based algorithm to the optimal solution of Problem \ref{Problem: Original UE and beam selection Problem}.}. 

Let  $\mcP_i:=\{ l \in \mcL| l_{\ta}=i, z_l>0 \}$ denote the set of active players at AP $i$. Let $\mcM_l^{\tg}:=\{ i' \in \mcA|{l'}_{\ta}=i', z_{l'}>0, l' \in \mcN_{l}^{\tg}\}$ denote the set of APs, at least one of whose active players, can create interference to player $l$. Designing the utilities of the players in an LIG is a crucial task. Now we provide our definition of utility; however, before providing the mathematical expression of the utility of a player in our LIG setting, we provide its interpretation through an example. We assume that all the active players at an AP transmit for equal fractions of time. Consider a network with $3$ APs and $9$ active players in which the APs corresponding to the players belonging to sets $\{1,2\}$, $\{3,4,5\}$, and $\{6,7,8,9\}$ are AP 1, AP 2, and AP 3, respectively. Suppose $i$, $j$, and $k$ are distinct and are players at different APs. Let $R(i,\{j,k\})$ denote the rate of the UE corresponding to player $i$ when players $i$, $j$, and $k$ concurrently transmit. For example, $R(8, \{1,4\})$ denotes the rate of the UE corresponding to player 8 when player 1 at AP 1, player 4 at AP 2, and player 8 at AP 3 concurrently transmit. In this example, we assume that each player affects the rates of all other players (which may not be the case in general).  Note that the fraction of time a player at AP 3  concurrently transmits with one of the players at AP 1 and one of the players at AP 2 is $1/24$. The utility of a player is the time average of the rates of the player over all possible sets of concurrent transmissions. For example, the utility of player 8 is $\frac{1}{24} \sum_{i \in \{1,2\}, j \in \{3,4,5\}} R(8,\{i,j\})$. Note that when only one player is active at each AP, then the utility of each of the players will be the same as its rate. Now we define the utility of a player in general. Let $\mcN_l^{\tut}:=\{ l' \in \mcL \setminus l| {l'}_{\ta} =i,  i \in \bigcup_{l'' \in \{ l \bigcup \mcN_{l}^{\tg} \} }  {l''}_a
\}$; it will soon be clear that $\mcN_l^{\tut}$ denotes the set of players whose strategy can affect the utility of player $l$. Let $V_l: \{\mcZ_l, \mcZ_{\mcN_l^{\tut}} \} \rightarrow \mcR$, where $\mcR$ is the set of real numbers, denote the utility of player $l$; it is defined as follows:
\begin{align*}
    V_l(z_l, z_{ \mcN_l^{\tut} } )= \frac{1}{\prod_{i \in \{ l_{\ta} \bigcup \mcM_{l}^g \} } k_i } \sum_{\mcP \in \times_{i \in \mcM_{l}^g} \mcP_i } R(l,\mcP),
\end{align*}
where $k_i=|\mcP_i|$, and 
\begin{align*}
    R(l,\mcP)=\begin{cases}
        0, & \\ \mbox{ if } z_l=0  \mbox{ or } l_{\tu}={l'}_{\tu} \text{ for some } l' \in \mcP,\\
        w_{l_{\tu}} B \log_2 \Bigg( 1+ \frac{S_{l_{\tb},l_{\tu},l_{\ta}}}{N_0+\sum_{l'\in \mcP}  S_{{l'}_b,l_{\tu},{l'}_{\ta}} } \Bigg),  & \\ \quad \mbox{otherwise}.
    \end{cases}
\end{align*}
In the above expressions, $R(l,\mcP)$ is the rate of the UE corresponding to player $l$ when player $l$ concurrently transmits with the players belonging to the set $\mcP$. Also, the utility, $V_l(z_l, z_{ \mcN_l^{\tut} })$, is the time average of the rates under all possible sets of concurrent transmissions involving player $l$.   

We define the payoff of player $l$ as the sum of the utilities of player $l$ and the players whose utility gets affected by the strategy of player $l$. Note that the payoff of player $l$ depends on the strategies of the players $\mcN_l^{\tp}:= \{ \bigcup_{l' \in \{ l \bigcup \mcN_l^{\ta} \bigcup \mcN_l^{\tu} \bigcup \mcN_l^{\tc} \}} \mcN_{l'}^{\tut} \}\setminus l  $. Let $Y_l: \{ \mcZ_l, \mcZ_{\mcN_l^{\tp}}  \} \rightarrow \mcR$ denote the payoff of player $l$. Then:
\begin{align*}
    Y_l(z_l, z_{\mcN_l^{\tp} })=\sum_{l' \in \{ l \bigcup\mcN_l^{\ta} \bigcup\mcN_l^{\tu} \bigcup\mcN_l^{\tc} \} }  V_{l'}(z_{l'}, z_{\mcN_{l'}^{\tut} }). 
\end{align*}

Let $\mathbf{G}:=[\mcL, \{ \mcZ_l \}_{l \in \mcL}, \{ Y_l \}_{l \in \mcL}]$ denote an LIG in strategic form with the set of players $\mcL$, set of strategies $\{ \mcZ_l \}_{l \in \mcL}$, and set of payoffs $\{ Y_l \}_{l \in \mcL}$.

\begin{definition}[Pure Strategy Nash Equilibrium (NE)] \label{Definition: GNE}
 A strategy profile $z^{\ast} \in \mcZ_{\mcL}$ for the game $\mathbf{G}$ is said to be a pure strategy NE if for all players $l \in \mcL$:
\begin{align*}
    Y_l({z_l^{\ast}}, {z_{-l}^{\ast}}) \ge  Y_l({z_l}, {z_{-l}^{\ast}}),  \quad \forall z_l \in \mcZ_l.
\end{align*}
\end{definition}

\begin{definition}[Potential Game] \label{Definition: Constrained Potential Game}
 Let $\Psi:\mcZ_{\mcL} \rightarrow \mcR$. Then, the function $\Psi$ is called a potential function for the LIG $\mathbf{G}$ if for all $l \in \mcL$ and $z_{-l} \in \mcZ_{-l}$:
\begin{align*}
    Y_l({z'}_l, z_{-l})-Y_l(z_l, z_{-l})=\Psi({z'}_l, z_{-l})-&\Psi(z_l, z_{-l}),\\
    &\forall z_l, {z'}_l \in \mcZ_l.
\end{align*}
Also, the game $\mathbf{G}$ is called a potential game.
\end{definition}

\subsection{Potential Game Characterization and Analysis of NE}
Let $\Psi(z_l, z_{-l}):=\sum_{l \in \mcL} V_l(z_l,z_{-l})$ denote the sum of utilities of all the players. It is easy to see that it is equal to the sum of weighted rates of all the UEs when each UE and each AP corresponds to at most one active player.  We establish the following results for the formulated LIG.
\begin{theorem} \label{Theorem: G is CPG}
    The game $\mathbf{G}$ is a potential game with potential function $\Psi$; also,  it has at least one pure strategy NE and the global maximizer of $\Psi$ is a pure strategy NE.
\end{theorem}

Note that in a potential game, the best response algorithm-- in which, each player chooses its strategy that maximizes its payoff given the strategies of the other players in a round-robin fashion-- converges with an exponential rate to a pure strategy NE of the game irrespective of the initial strategy~\cite{swenson2018best}. However, the obtained pure strategy NE may not correspond to the global optimum of the potential function. In the following subsection, we provide an algorithm that achieves a global maximum of $\Psi$.

\subsection{Proposed LIG-based UE and Beam Selection Algorithm}
In Algorithm \ref{Algorithm: LIG based UE and Beam Selection}, we provide an algorithm for UE and beam selection in mmWave networks. The algorithm is similar to the concurrent spatial adaptive play (CSAP)  algorithm proposed in~\cite{xu2011opportunistic} for achieving an optimal strategy for LIGs. Algorithm \ref{Algorithm: LIG based UE and Beam Selection} is written using notations and definitions some of which are not defined earlier. We now describe those notations and definitions.

\begin{algorithm} 
    \algsetup{linenosize=\normalsize}
	\caption{LIG-based UE and Beam Selection} 
	\begin{algorithmic}[1] \label{Algorithm: LIG based UE and Beam Selection}
	\STATE \textbf{Input:} $S_{b,u,a}, \forall b \in \mcB, u \in \mcU,  a \in \mcA,$
	
	\STATE \textbf{Output:}  $z_l, \forall l \in \mcL$ // Strategy for each Player
	\STATE Initialization: $i=0, z_l(0)=0, \forall l \in \mcL$ 
 	\WHILE{$i \le I_0$} 
	\STATE Choose an independent set of players $\Hat{\mcL} (i) \in \mbL$
        \FOR{ each $l \in \Hat{\mcL} (i)$}
        \STATE Compute $Y_l\big(z_l, z_{\mcN_l^{\tp}}(i-1)\big), \forall z_l \in \mcZ_l$ // Payoffs of all strategies of player $l$
        \STATE Compute $\Pr(z_l(i)=z_l)=\frac{e^{\beta Y_l\big(z_l, z_{\mcN_l^{\tp}}(i-1)\big) }}{\sum_{{z'}_l \in \mcZ_l } e^{\beta Y_l\big({z'}_l, z_{\mcN_l^{\tp}} (i-1)\big) } }, \forall z_l \in \mcZ_l$ // Probability that $z_l$ will be the updated strategy for player $l$
        \STATE Let $\mu_l(i)=\{\Pr(z_l(i)=z_l)|z_l \in \mcZ_l\}$ 
        \STATE Update $z_l(i)$ using the probability distribution $\mu_l(i)$
        \ENDFOR
        \STATE $i=i+1$
	\ENDWHILE
        \STATE Output $z_l=z_l(I_0), \, \forall l \in \mcL$
	\end{algorithmic} 
\end{algorithm}

\begin{definition}[Mutually Independent Set of Players]
    A set $\Hat{\mcL}\subset \mcL$ is said to be a mutually independent set of players for the game $\mathbf{G}$ if $\forall l, l' \in \Hat{\mcL}, \, \{ l \bigcup \mcN_l^{\tp} \}\bigcap \{ l' \bigcup \mcN_{l'}^{\tp} \}=\emptyset$.
\end{definition}
Note that the number of mutually independent sets of players can be more than 1.  Let $i \in \{0,1,2, \cdots \}$ denote the iteration number for strategy update of players. Let $\Hat{\mcL} (i)$ (respectively, $\mbL$) denote a mutually independent set of players chosen for strategy update in the $i$th iteration (respectively, set of all mutually independent sets of players). The mutually independent set $\Hat{\mcL} (i) \in \mbL, \, \forall i \in \{0, 1, 2, \cdots\}$, and it is chosen randomly from the set $\mbL$ with uniform probabilities $\frac{1}{|\mbL|}$. The number of mutually independent sets of players, i.e., $|\mbL|$ depends on the scale of the network in terms of the number of APs and UEs, transmit power level at APs, the density of APs and UEs, etc. Let $z_l(i)$ (respectively, $z_{\mcN_l^{\tp}}(i)$) denote the strategy of player $l \in \mcL$ (respectively, strategy profile of players belonging to set $\mcN_l^{\tp}$) after the $i$th strategy update.  Let
\begin{align}
    \Pr(z_l(i)=z_l):=\frac{e^{\beta Y_l\big(z_l, z_{\mcN_l^{\tp}}(i-1)\big) }}{\sum_{{z'}_l \in \mcZ_l} e^{\beta Y_l\big({z'}_l, z_{\mcN_l^{\tp}} (i-1)\big) } }, \forall z_l \in \mcZ_l \label{Eq: LIG, probability that z_l will be updated strategy}
\end{align}
 be the probability with which a strategy $z_l \in \mcZ_l$ becomes the updated strategy of player $l$, where $\beta > 0$ is  a parameter. Let $\mu_l(i):=\{\Pr(z_l(i)=z_l)|z_l \in \mcZ_l\}$ denote the probability distribution using which the strategy of player $l$ is updated in the $i$th update. Algorithm \ref{Algorithm: LIG based UE and Beam Selection} starts with the initial strategy $z_l(0)=0, \forall l \in \mcL$. In each iteration $i$, Algorithm \ref{Algorithm: LIG based UE and Beam Selection} updates the strategies $z_l(i)$ of players $l \in \Hat{\mcL}(i)$, where $\Hat{\mcL}(i) \in \mbL$, by computing-- first, the payoffs of the players for all their strategies, and second, the probability distribution for each player over its strategies. 

Let $I_0$ denote the number of times for which Algorithm \ref{Algorithm: LIG based UE and Beam Selection} updates the strategies of players. Note that Algorithm \ref{Algorithm: LIG based UE and Beam Selection} can be terminated in two ways: (i) the first, as done in the above pseudocode for Algorithm \ref{Algorithm: LIG based UE and Beam Selection}, where it gets terminated after some fixed number of iterations, $I_0$, and  (ii) the second, when for each player, the  probability of one of its strategies computed using (\ref{Eq: LIG, probability that z_l will be updated strategy}), falls within some $\epsilon \in (0,1)$ distance from 1.  

\subsection{Optimality of LIG-based UE and Beam Selection}
In this subsection, we show that our proposed Algorithm \ref{Algorithm: LIG based UE and Beam Selection} achieves the optimal UE and beam selection for Problem \ref{Problem: Original UE and beam selection Problem}. Also, we provide an upper bound on the expected number of iterations required by Algorithm \ref{Algorithm: LIG based UE and Beam Selection} to converge to the global optimum for the first time. Some of our results are similar to those in~\cite{frigessi1993convergence,roberts1994simple,young1998individual,marden2009cooperative,zhang2011weighted,xu2011opportunistic,zheng2015optimal,liu2018decentralized}. 
\begin{lemma} \label{Lemma: stationary distribution LIG}
     Algorithm \ref{Algorithm: LIG based UE and Beam Selection}  obtains the following unique stationary distribution:
    \begin{align*}
        \pi (z)=\frac{e^{\beta \Psi(z)}}{\sum_{z'\in \mcZ_{\mcL}}e^{\beta \Psi(z')}}, \quad \forall z \in \mcZ_{\mcL},
    \end{align*}
   where $\mcZ_\mcL$ is the strategy space of the players and $\Psi$ is the potential function of the game $\mathbf{G}$.
\end{lemma}
Lemma \ref{Lemma: stationary distribution LIG} states that Algorithm \ref{Algorithm: LIG based UE and Beam Selection} achieves a unique stationary distribution over the strategy space of the players.
\begin{theorem} \label{Theorem: LIG attain global opt. of PSI}
    Let $\mcZ_{\mcL}^{\topt}:=\argmax_{z \in \mcZ_{\mcL}} \Psi(z)$ be the set of the strategy profiles of the players that achieve the global maximum of $\Psi$. Then:  
    \[
    \lim_{\beta \uparrow \infty } \pi (z)=        \begin{cases}
           \frac{1}{|\mcZ_{\mcL}^{\topt}|},& \forall z  \in \mcZ_{\mcL}^{\topt}, \\
           0, & \mbox{otherwise}.
        \end{cases}
    \]
\end{theorem}

Theorem \ref{Theorem: LIG attain global opt. of PSI} states that Algorithm \ref{Algorithm: LIG based UE and Beam Selection} achieves the global optimum of $\Psi$. However, recall that we want to show that Algorithm \ref{Algorithm: LIG based UE and Beam Selection} achieves the global optimum of Problem \ref{Problem: Original UE and beam selection Problem}. 

Let $\mcZ_{\mcL}^{\tc}:=\{ z \in \mcZ_{\mcL}| |\{ l \in \mcL| l_{\ta}=i, z_l>0 \}| \le 1, \forall i \in \mcA, \mbox{ and } |\{ l \in \mcL| l_{\tu}=u, z_l>0 \}| \le 1, \forall u \in \mcU  \}$ denote the set of the strategies of players that satisfy the constraints-- at most one player remains on at any AP and each UE gets served by at most one AP-- of the original  UE and beam selection problem, Problem \ref{Problem: Original UE and beam selection Problem}.
\begin{lemma} \label{Lemma: Strategy Satisfy Constraints}
If $\Hat{z} \in \{ \mcZ_{\mcL} \setminus \mcZ_{\mcL}^{\tc}\}$, then there exists a $\Tilde{z} \in \mcZ_{\mcL}$ satisfying $\Tilde{z}_{l^*}=0$ and $\Tilde{z}_l=\Hat{z}_l, \forall l \in \mcL \setminus l^*$ for some $l^*$ having $\Hat{z}_{l^*}>0$ such that $\Psi(\Tilde{z}) \ge \Psi(\Hat{z})$.
\end{lemma}
Lemma \ref{Lemma: Strategy Satisfy Constraints} states that the sum of utilities of the players can be either improved or kept unchanged by switching off one of the players when the strategy profile of the players does not satisfy the constraints of Problem \ref{Problem: Original UE and beam selection Problem}.

\begin{theorem}\label{Theorem: LIG Algo achieve optimal UE and beam selection}
For sufficiently large $\beta$, the global maximum of $\Psi$ obtained by Algorithm \ref{Algorithm: LIG based UE and Beam Selection}  corresponds to the optimal UE and beam selection of Problem \ref{Problem: Original UE and beam selection Problem}.
\end{theorem}
Theorem \ref{Theorem: LIG Algo achieve optimal UE and beam selection} states that the \emph{proposed LIG-based UE and beam selection algorithm achieves the optimal UE and beam selection for Problem \ref{Problem: Original UE and beam selection Problem}}.

In the following, we provide an upper bound on the expected number of iterations required by Algorithm \ref{Algorithm: LIG based UE and Beam Selection} to reach the optimal UE and beam selection for Problem \ref{Problem: Original UE and beam selection Problem}. We provide this bound by obtaining an upper bound on the expected number of iterations required by an algorithm, say \algDLIG, which is similar to Algorithm \ref{Algorithm: LIG based UE and Beam Selection}, with the difference that in each iteration, the algorithm updates the strategy of exactly one player, which is chosen uniformly at random from the set of all players $\mcL$.
Let $Z(i)$ be the strategy profile of the players chosen by {\algDLIG} in the $i$th iteration.  Note that $\{Z(i)\}_{i \ge 0}$ is a Markov chain over the strategy space $\mcZ_{\mcL}$. We construct an auxiliary  Markov chain $\{ \widetilde{Z}(i) \}_{i \ge 0}$ from $\{Z(i)\}_{i \ge 0}$  such that once the auxiliary chain enters into one of the states in $\mcZ_{\mcL}^{\topt}$, it remains there. That is, $\widetilde{Z}(i) \in \mcZ_{\mcL}^{\topt} \implies\widetilde{Z}(i') \in \mcZ_{\mcL}^{\topt}, \, \forall i' \ge i$ and $\widetilde{Z}(i)=Z(i), \, \forall i \le \argmin\{Z(i') \in \mcZ_{\mcL}^{\topt}|i' \ge 0\}$. To obtain an upper bound on the expected number of strategy updates required by {\algDLIG} to reach the set of globally optimal strategies, we use the concept of an absorbing Markov chain.
\begin{definition}[Absorbing Markov Chain~\cite{ermon2014designing}]
    Let $\{ Z(i) \}_{i \ge 0}$ be a Markov chain over $\mcZ$. A state $z \in \mcZ$ is said to be an absorbing state if the chain leaves state $z$ with zero probability once it enters into it. If $\{Z(i)\}_{i \ge 0}$ has at least one absorbing state, then it is an absorbing Markov chain.
\end{definition}

Note that $\{ \widetilde{Z}(i) \}_{i \ge 0}$ is an absorbing Markov chain with set of absorbing states $\mcZ_{\mcL}^{\topt}$.  Let $k_z:=\min \{ i \in \mathbb{N}|\widetilde{Z}(i) \in \mcZ_{\mcL}^{\topt}, \widetilde{Z}(0)=z, z \in \mcZ_{\mcL}\}$ denote the minimum number of steps required to reach the set of absorbing states when only one player changes its strategy in each step and the chain starts from state $z$. Note that the minimum number of steps required by Algorithm \ref{Algorithm: LIG based UE and Beam Selection} to reach the set of absorbing states when the chain starts in $z$ will be less than  $k_z$  since multiple players can change their strategies in each step. Let $D:=\max\{ 
k_z|z \in \mcZ_{\mcL} \}$ denote the minimum number of steps required by {\algDLIG} to reach the set of absorbing states from any state in the strategy space. Let $\nu_i:=\Pr(\widetilde{Z}(i) \in \mcZ_{\mcL}^{\topt})$ denote the probability of being in one of the absorbing states in iteration $i$. It is easy to see that $\nu_i$ is non-decreasing in $i$. Let $\eta:=\min_{\Tilde{z},\Hat{z} \in \mathcal{Z_{\mcL} }} \Pr(\widetilde{Z}(i+1)=\Tilde{z}|\widetilde{Z}(i)=\Hat{z})$ denote the minimum transition probability from a state $\Hat{z}$ to another state $\Tilde{z}$ in the strategy space when only one player is allowed to change its strategy in a step. Then it is easy to see that: 
\begin{align*}
    \eta=\min_{z \in \mcZ_{\mcL}}\bigg( \min_{l \in \mcL} \bigg( \min_{z_l \in \mcZ_l} \frac{e^{\beta Y_l \big(z_l, z_{\mcN_l^{\tp}}\big)}}{\sum_{{z'}_l \in \mcZ_l} e^{\beta Y_l \big({z'}_l, z_{\mcN_l^{\tp}}\big)} } \bigg) \bigg).
\end{align*}

\begin{theorem} \label{Theorem: No. of iter. to reach optimum}
    The expected number of iterations required by Algorithm \ref{Algorithm: LIG based UE and Beam Selection} to reach the global optimum for the first time is upper bounded by: 
    \begin{align*}
        \frac{D (1-\nu_0)}{\eta^D}.
    \end{align*}
\end{theorem}

\section{Need for Design of New UE and Beam Selection Algorithm} \label{Section: Requirement for design of new algorithm}
\begin{figure}
	\centering
	\begin{subfigure}{.49\textwidth}
              \hspace*{-0.2cm}
              \begin{tikzpicture}[scale=1]
                \begin{axis}[
                  scale=1.45,
                  ybar=2pt,
                  symbolic x coords={A, B, C, D},
                  xtick=data,
                  xticklabels={InH, UMi, UMa, RMa},
                  xticklabel style={font={\tiny \boldmath}, rotate=0},
                  minor x tick num=1,
                  minor grid style={dashed},
                  major x tick style={draw=none},
                  minor x tick style={draw=none},
                  enlarge x limits=0.2,
                   ymax=4.2,
                  yticklabel style={font=\tiny \boldmath },
                  ylabel={Throughput per UE (Gbps)},
                  label style={font={\tiny \bfseries}},
                  ylabel style={yshift=-5pt},
                  minor y tick num=1,
                  minor grid style={dashed},
                  enlarge y limits=0.2,
                  major tick length=2pt,
                  minor tick length=1pt,
                  axis line style = thick,
                  legend cell align=left,
                  legend columns=6,
                 legend style={thick, font={\tiny}, draw=black, fill=white, at={(0.5,1.25)}},
                 width=\textwidth,
                 bar width=0.078cm, 
                 nodes near coords, 
                 visualization depends on={\thisrow{error} \as \myoffset},
                 every node near coord/.append style={yshift=\myoffset*0.1 cm, black, font={\tiny \boldmath},
                  rotate=90, anchor=west}
                  ]  
            
              \addplot [cyan, ultra thin, fill=cyan, error bars/.cd, y dir=both, y explicit, error bar style={red, very thick}] table [x=position, y=value, y error=error] {
                    value   position    error
                    2.2186  A           0.0136
                    3.2402  B           0.0282
                    3.2321  C           0.0356
                    3.2800  D           0.0305        
                };
              \addplot [black, ultra thin, fill=black, error bars/.cd, y dir=both, y explicit, error bar style={red, very thick}] table [x=position, y=value, y error=error] {
                    value   position    error
                    2.2186  A           0.0136
                    3.2402  B           0.0282
                    3.2321  C           0.0356
                    3.2800  D           0.0305        
                }; 
            \addplot [red, ultra thin, fill=red, error bars/.cd, y dir=both, y explicit,  error bar style={black, very thick}] table [x=position, y=value, y error=error] {
                    value   position    error
                    2.2186  A           0.0111
                    3.2381  B           0.0274
                    3.2198  C           0.0375
                    3.2772  D           0.0293
                };
            \addplot [blue, ultra thin, pattern=horizontal lines, 
            pattern color=blue,  error bars/.cd, y dir=both, y explicit, error bar style={blue, very thick}] table [x=position, y=value, y error=error] {
                    value   position    error
                    1.5100  A           0.0219
                    2.3649  B           0.0401
                    2.2448  C           0.0351
                    2.4342  D           0.0408
                };
            \addplot [green, ultra thin, pattern=crosshatch, 
            pattern color=green, error bars/.cd, y dir=both, y explicit, error bar style={green, very thick}] table [x=position, y=value, y error=error] {
                    value   position    error
                    1.1059  A           0.0179
                    0.9907  B           0.0376
                    0.9750  C           0.0325
                    1.1352  D           0.0407
                };
            \addplot [magenta, ultra thin, pattern= grid, 
            pattern color=magenta, error bars/.cd, y dir=both, y explicit, error bar style={magenta, very thick}] table [x=position, y=value, y error=error] {
                    value   position    error
                    1.0060  A           0.0214
                    0.9494  B           0.0264
                    0.9036  C           0.0229
                    1.0551  D           0.0319
                };
               \legend{Exhaustive Search}
                \end{axis}
              \end{tikzpicture}
		\caption{}
		\label{Fig: Motivation for new heuristics}
	\end{subfigure}
	\begin{subfigure}{.49\textwidth}
            \hspace*{-1.75cm}
            \begin{tikzpicture}[scale=1]
                \begin{axis}[
                  scale=1.45,
                  ybar=1.5pt,
                  symbolic x coords={A, B, C, D},
                  xtick=data,
                  xticklabels={InH, UMi, UMa, RMa},
                  xticklabel style={font={\tiny \boldmath}, rotate=0},
                  minor x tick num=1,
                  minor grid style={dashed},
                  major x tick style={draw=none},
                  minor x tick style={draw=none},
                  enlarge x limits=0.2,
                   ymax=4.5,
                  yticklabel style={font=\tiny \boldmath},
                  label style={font={\tiny \bfseries}},
                  ylabel style={yshift=-5pt},
                  minor y tick num=1,
                  minor grid style={dashed},
                  enlarge y limits=0.2,
                  major tick length=2pt,
                  minor tick length=1pt,
                  axis line style = thick,
                  legend cell align=left,
                  legend columns=7,
                 legend style={thick, font={\tiny}, draw=black, fill=white, at={(1.44,1.25)},  xshift=-50pt, yshift=0pt},
                 width=\textwidth,
                 bar width=0.079cm, 
                 nodes near coords, 
                 visualization depends on={\thisrow{error} \as \myoffset},
                 every node near coord/.append style={yshift=\myoffset*0.1 cm, black, font={\tiny \boldmath},
                  rotate=90, anchor=west}
                  ]  
            
              \addplot [black, ultra thin, fill=black, error bars/.cd, y dir=both, y explicit, error bar style={red, very thick}] table [x=position, y=value, y error=error] {
                    value   position    error
                    2.7000  A           0.0136
                    3.8850  B           0.0282
                    3.9748  C           0.0356
                    4.0681  D           0.0305        
                };
              \addplot [red, ultra thin, fill=red, error bars/.cd, y dir=both, y explicit, error bar style={black, very thick}] table [x=position, y=value, y error=error] {
                    value   position    error
                    2.7000  A           0.0136
                    3.8850  B           0.0282
                    3.9748  C           0.0356
                    4.0681  D           0.0305        
                };
              \addplot [black, ultra thin, pattern=north east lines, 
            pattern color=black, error bars/.cd, y dir=both, y explicit, error bar style={black, very thick}] table [x=position, y=value, y error=error] {
                    value   position    error
                    2.3864  A           0.0136
                    3.7624  B           0.0282
                    3.8245  C           0.0356
                    3.9856  D           0.0305        
                }; 
            \addplot [red, ultra thin, pattern=north west lines, 
            pattern color=red, error bars/.cd, y dir=both, y explicit,  error bar style={red, very thick}] table [x=position, y=value, y error=error] {
                    value   position    error
                    2.5747  A           0.0111
                    3.7299  B           0.0274
                    3.9748  C           0.0375
                    4.0681  D           0.0293
                };
            \addplot [blue, ultra thin, pattern=horizontal lines, 
            pattern color=blue,  error bars/.cd, y dir=both, y explicit, error bar style={blue, very thick}] table [x=position, y=value, y error=error] {
                    value   position    error
                    1.8638  A           0.0219
                    2.4522  B           0.0401
                    2.9534  C           0.0351
                    3.2743  D           0.0408
                };
            \addplot [green, ultra thin, pattern=crosshatch, 
            pattern color=green, error bars/.cd, y dir=both, y explicit, error bar style={green, very thick}] table [x=position, y=value, y error=error] {
                    value   position    error
                    1.1059  A           0.0179
                    0.9907  B           0.0376
                    1.5764  C           0.0325
                    1.0819  D           0.0407
                };
            \addplot [magenta, ultra thin, pattern= grid, 
            pattern color=magenta, error bars/.cd, y dir=both, y explicit, error bar style={magenta, very thick}] table [x=position, y=value, y error=error] {
                    value   position    error
                    1.2142  A           0.0214
                    0.9494  B           0.0264
                    0.9036  C           0.0229
                    1.2993  D           0.0319
                };
                \legend{MCMC, LIG, NGUB1, NGUB2, PIA, MDSR,BounceNet}
                \end{axis}
              \end{tikzpicture}
		\caption{}
		\label{Fig: vScen}
	\end{subfigure}
	\caption{The plots show a performance comparison in terms of the per-user average throughput metric under different algorithms for different mmWave network scenarios. The common parameters used in Fig.~\ref{Fig: Motivation for new heuristics} and Fig.~\ref{Fig: vScen} are operating frequency= 60 GHz and number of schedules per slot (SPS)=1. The other parameters used in Fig.~\ref{Fig: Motivation for new heuristics} (respectively, Fig.~\ref{Fig: vScen}) are $N_A$ =4, $N_U$ = 10 (respectively, $N_A$ =9, $N_U$ = 25).}
	\label{fig3}
\end{figure}
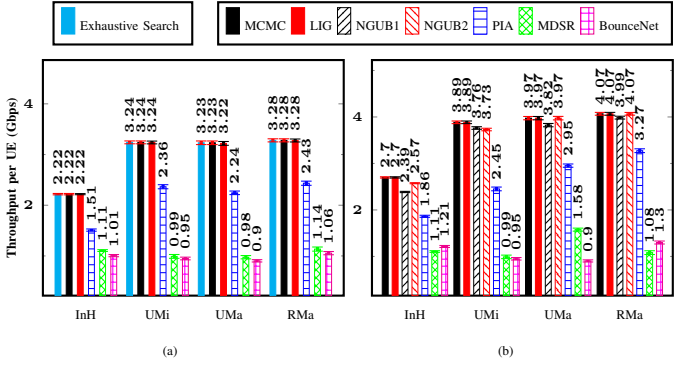

In prior work, several algorithms have been proposed for UE and beam selection in mmWave networks. These algorithms have been designed for some mmWave network scenarios based on insights obtained either from simulations or experimentation. The most relevant and recent algorithms for UE and beam selection in prior works are in~\cite{wei2017pose,jog2019many,yang2021mdsr} and are described briefly in~\cite{singh2023ngub}. All of these works show that their proposed algorithm outperforms some algorithms  from prior work; however, none of them say anything about how closely the algorithm performs compared to an optimal algorithm. Recall that in Section \ref{Section: MCMC based UE and Beam Selection} (respectively, Section \ref{Section: LIG based UE and Beam Selection}), we proposed an MCMC-based (respectively, LIG-based) algorithm to find asymptotically optimal UE and beam selections. These algorithms can be used to evaluate the performance of a designed heuristic by finding the gap in performance corresponding to the solution obtained by the heuristic and the solution provided by the asymptotic algorithm. In Figure \ref{Fig: Motivation for new heuristics}, we show that the MCMC-based and LIG-based algorithms provide the same per-user throughput as obtained by the exhaustive search-based UE and beam selection algorithm in different mmWave network scenarios--Indoor Hall (InH), Urban Micro (UMi), Urban Macro (UMa), and Rural Macro (RMa) (details of the simulation setup are described in Section \ref{Section: Performance Evaluation}).  Also, the figure shows that the most relevant algorithms in prior work-- PIA~\cite{wei2017pose}, MDSR~\cite{yang2021mdsr}, and BounceNet~\cite{jog2019many} underperform by a significant margin with respect to the proposed MCMC-based and LIG-based asymptotic algorithms.

Note that all the above mentioned algorithms in prior work first perform preliminary UE association and then do joint UE and beam selection using the preliminary association, which might lead to poor UE and beam selection. 
In Section \ref{Section: Proposed Algorithm}, we propose two fast greedy algorithms-- NGUB1 and NGUB2, which perform joint UE and beam selection in a single step and outperform the above algorithms from prior work.
\section{Proposed Algorithms} \label{Section: Proposed Algorithm}
We propose two fast greedy algorithms, NGUB1 and NGUB2, for joint UE and beam selection in a mmWave network. 
\subsection{NGUB1}
Let $\Bar{\mcA}^i$  denote the set of APs that already got a $({\tu}_a, {\tb}_a)$ pair till the beginning of the $(i+1)$th iteration. Let $\Bar{\mcU}^i$ denote the set of UEs that are already associated with some $AP$ in $\Bar{\mcA}^i$ till the beginning of the $(i+1)$th iteration. Note that $\Bar{\mcA}^0=\emptyset$ and $\Bar{\mcU}^0=\emptyset$. Let ${\tb}_a^u$ denote the best beam at AP $a$ for UE $u$ based on RSS. The algorithm runs in two steps-- the first, \textit{Initial Selection}, and the second, \textit{Improvement over the Initial Selection}.
\subsubsection{Initial Selection}
In the $i$th iteration, the algorithm selects an AP $a$ and  a UE ${\tu}_a$ such that:
\begin{align}
         <{\tu}_a,a>=&\argmax_{(u,a)|a \in \mcA \backslash \bar{\mcA}^{i-1}, u \in \mcU \backslash \bar{\mcU}^{i-1}}  w_u B \times \label{Eq: Matrix M computation NG1}\\
         &\log_2 \Bigg( 1 + \frac{S_{{\tb}_a^u,u,a}}{N_0+\sum_{a' \in \bar{\mcA}^{i-1}} S_{b_{a'}^{u'},u,a'}} \Bigg). \nonumber
     \end{align}
   In (\ref{Eq: Matrix M computation NG1}), $u'$ is the UE assigned to the AP $a' \in \bar{\mcA}^{i-1}$. Note that the above process of initial selection is complete after $\min (N_A, N_U)$ iterations. Also, the computation required in each iteration decreases with the iteration number.
\subsubsection{Improvement over Initial Selection}
In this step, in each iteration, an AP is selected in a round-robin fashion. A new UE from the set of free UEs is assigned to the selected AP if the new UE maximizes, across all the free UEs, the weighted sum rate, and the new weighted sum rate is larger than the current weighted sum rate; otherwise, the UE does not change at the selected AP. Note that the UEs at the other APs remain the same as in the current selection. Let ${\tu}_a^{\tc}$ (respectively, ${\tu}_a^{\tn}$)  denote the current UE (respectively, new UE) for AP $a$. Let $\mcU_a^{\tc}:=\mcU \backslash \{ u|u=u_{a'}^{\tc} \mbox{ for some }a' \in \mcA \mbox{ and } a' \neq a \}$ denote the set of free UEs that can replace ${\tu}_a^{\tc}$ at AP $a$. Let $r_{u,{\tu}_a^{\tc}}^{\tn}$ denote the new weighted sum rate when ${\tu}_a^{\tc}$ is replaced by UE $u$. Then,
\begin{align*}
    {\tu}_a^{\tn}=\argmax_{u \in \mcU_a^{\tc}\bigcup \{{\tu}_a^{\tc}\}}r_{u,{\tu}_a^{\tc}}^{\tn}.
\end{align*}
 This step can be terminated in two ways-- (i) after a fixed number of iterations and (ii) after the increment in the weighted sum rate becomes less than a threshold value after one complete round over APs. In our simulations, we use the first stopping criterion.

\subsection{NGUB2}
In this algorithm, a  \textit{Modified Initial Selection} step-- similar to \textit{Initial Selection} of the \textit{NGUB1} algorithm with some modification-- is followed some fixed number of times, let us say $J$ times, where $J$ can depend on $N_A$ and/ or $N_U$. Each time the step is followed, it gives a set of pairs $\{({\tb}_a, {\tu}_a), a \in \mcA\}$. Let $\Breve{\mcA}^j$ (respectively, $R_j$) denote the set of the beam and UE pairs (respectively, the weighted sum rate of UEs corresponding to $\Breve{\mcA}^j$) when the \textit{Modified Initial Selection} step is run for the $j$th time. The algorithm stores   $\Breve{\mcA}^j$ along with $R_j$ in memory and after the completion of $J$ runs of the \textit{Modified Initial Selection} step, the algorithm chooses $\Breve{\mcA}^j$ for scheduling if $j=\argmax_{i\in \{1,2,\cdots,J\}} R_i$. 

In the $i$th iteration of a run of the \textit{Modified Initial Selection} step, the algorithm selects an AP $a \in \bar{\mcA}^{i-1}$ uniformly at random and a UE ${\tu}_a$ such that:
\begin{align*}
         {\tu}_a=&\argmax_{u| u \in \mcU 
\backslash \bar{\mcU}^{i-1}}  w_u B \times \\
         &\log_2 \Bigg( 1 + \frac{S_{{\tb}_a^u,u,a}}{N_0+\sum_{a' \in \bar{\mcA}^{i-1}} S_{b_{a'}^{u'},u,a'}} \Bigg).
     \end{align*}
   Note that the above process of selection is complete after $\min (N_A, N_U)$ iterations. Also, the computation required in each iteration decreases with the iteration number.

\subsection{Difference between Proposed Algorithms and the Algorithms in Prior Work}
The algorithms proposed in~\cite{wei2017pose,jog2019many,yang2021mdsr} perform the task of UE and beam selection in mmWave networks in two steps. In the first step, they select mutually exclusive subsets of UEs for each AP. In the second step, they select a UE from each subset obtained in the first step. Also, in the second step, they select a beam at each AP for the selected UE. Note that this two-step process may not provide an optimal set of UE and beam pairs for each AP. In contrast, our proposed algorithms, NGUB1 and NGUB2, jointly select UE and beam for each AP. Also, note that our proposed algorithms do not restrict themselves to a subset of UEs during UE selection. Intuitively, due to the above reasons, our proposed algorithms outperform those proposed in~\cite {wei2017pose,jog2019many,yang2021mdsr}, as demonstrated by the simulation results in Section \ref{Section: Performance Evaluation}. 

\section{Simulations} \label{Section: Performance Evaluation}
In this section, we compare the performance of the most relevant algorithms proposed in prior works with that of our proposed asymptotically optimal  MCMC-based and LIG-based algorithms, and proposed novel greedy algorithms, NGUB1 and NGUB2, for UE and beam selection in mmWave networks via simulations. To show the robustness of our algorithms, we consider different mmWave networks by varying the network parameters $N_A$, $N_U$, operating frequency, scenarios,  and the number of schedules per slot (SPS). We consider $(N_A,N_U)=(4,10), (9,25), (16,40)$,  frequency=28 GHz, 60 GHz, 73 GHz, scenarios= InH, UMi, UMa, RMa, and number of SPS=1, 5, 10. For the cases with $N_A=4$,  $N_A=9$, and $N_A=16$,  we placed APs in the form of a regular square grid of 4 nodes, 9 nodes, and 16 nodes, respectively. Ranges of values of parameters such as distances among APs, power at APs, speeds of UEs, azimuthal and elevation beam widths, etc., for different mmWave network scenarios are provided in~\cite{samimi20163d,ju20203d,ju2021millimeter}; in our simulations, we use values from these ranges.  We consider 50 m, 100 m, 200 m, and 500 m as the edge length of the grid for scenarios InH, UMi, UMa, and RMa, respectively. UEs move only at the beginning of a slot and follow a random walk process. Each AP has 36 beams, each beam having a beam width of 20 degrees in both azimuthal and elevation directions.  For all simulations, we consider the number of slots equal to 2000. We use the three-dimensional statistical channel model in~\cite{samimi20163d} for RSS generation and choose the parameters in the model as in~\cite{samimi20163d,ju20203d,ju2021millimeter}. Note that the channel model in~\cite{samimi20163d} is based on extensive field measurements conducted in different mmWave network scenarios and the RSSs generated by the model are similar to the original measurements in the statistical sense.  Let $N^{\ts}$ and $N^{\tsr}$ be the number of slots in each simulation run and the number of simulation runs, respectively. Recall that $K$ is the number of SPS. Let $R_{i,t,k}^u$  be the rate of UE $u$ in the $k$th schedule of the $t$th slot of the $i$th run. Let $\tau_i^u:=\frac{1}{K N^{\ts}} \sum_{t=1}^{N^{\ts}}\sum_{k=1}^{K}R_{i,t,k}^u$ (respectively, $\tau_i:=\frac{1}{K N^{\ts}N_U} \sum_{u=1}^{N_U} \sum_{t=1}^{N^{\ts}}\sum_{k=1}^{K}R_{i,t,k}^u$) be the throughput of UE $u$ (respectively, per user throughput of UEs) in the $i$th run of a simulation.  Let $\mu_u:=\frac{1}{N^{\tsr}}\sum_{i=1}^{N^{\tsr}} \tau_i^u$ (respectively, $\mu:=\frac{1}{N^{\tsr}}\sum_{i=1}^{N^{\tsr}} \tau_i$) and $\sigma_u:=\sqrt{\frac{1}{N^{\tsr}} \sum_{i=1}^{N^{\tsr}} (\tau_i^u-\mu_u)^2}$ (respectively, $\sigma:=\sqrt{\frac{1}{N^{\tsr}} \sum_{i=1}^{N^{\tsr}} (\tau_i-\mu)^2}$) denote the average throughput of UE $u$ (respectively, average per user throughput) and the standard deviation (SD) of throughput of UE $u$ (respectively, SD of per user throughput), respectively. We define the Jain's fairness index (JFI) as follows~\cite{jain1984quantitative}:
\begin{align*}
    \mathbf{JFI}=\frac{(\sum_{u=1}^{N_U} \mu_u)^2}{N_U \sum_{u=1}^{N_U} \mu_u^2 }.
\end{align*}
The value of $\mathbf{JFI}$ lies in $[0,1]$. Also, it increases with the degree of fairness of the distribution of the throughput of different users~\cite{jain1984quantitative}.    We use the per-user average throughput and the JFI as performance metrics. In particular, we have plotted the mean and the SD of the per-user throughput and the JFI obtained from 20 runs of the simulation for each algorithm and for each mmWave network scenario. 

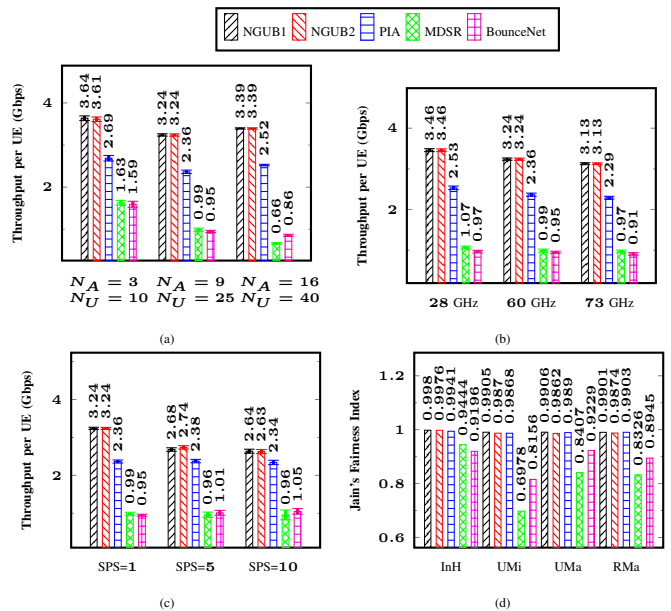
\begin{figure}
	\centering
	\begin{subfigure}{.49\textwidth}
		\centering
                \begin{tikzpicture}[scale=1]
                    \begin{axis}[scale=1.2,
                      ybar=2pt,
                      symbolic x coords={A, B, C},
                      xtick=data,
                      xticklabels={ 
                      \begin{tabular}{@{}l@{}} $N_A=3$ \\$N_U=10$ \end{tabular}, \hspace*{1em}
                      \begin{tabular}{@{}l@{}}$N_A=9$ \\$N_U=25$ \end{tabular}, \hspace*{2em}
                     \begin{tabular}{@{}l@{}}$N_A=16$ \\$N_U=40$ \end{tabular},
                      },
                      xticklabel style={font={\tiny \boldmath}, rotate=0},
                      minor x tick num=1,
                      minor grid style={dashed},
                      major x tick style={draw=none},
                      minor x tick style={draw=none},
                      enlarge x limits=0.3,
                       ymax=4.5,
                      yticklabel style={font=\tiny \boldmath },
                      ylabel={Throughput per UE (Gbps)},
                      label style={font={\tiny \bfseries}},
                      minor y tick num=1,
                      minor grid style={dashed},
                      enlarge y limits=0.1,
                      major tick length=2pt,
                      minor tick length=1pt,
                      axis line style = thick,
                      legend cell align=left,
                      legend columns=5,
                     legend style={thick, font={\tiny}, draw=black, fill=white, at={(2.0,1.3)}},
                     width=\textwidth,
                     bar width=0.09cm, 
                     nodes near coords, 
                     visualization depends on={\thisrow{error} \as \myoffset},
                     every node near coord/.append style={yshift=\myoffset*0.1 cm, black, font={\tiny \boldmath},
                      rotate=90, anchor=west}
                      ]  
                  \addplot [black, ultra thin, pattern=north east lines, 
                pattern color=black, error bars/.cd, y dir=both, y explicit, error bar style={black, very thick}] table [x=position, y=value, y error=error] {
                        value   position    error
                        3.6371  A           0.0555
                        3.2402  B           0.0282
                        3.3943  C           0.0151
                    }; 
                \addplot [red, ultra thin, pattern=north west lines, 
                pattern color=red, error bars/.cd, y dir=both, y explicit,  error bar style={red, very thick}] table [x=position, y=value, y error=error] {
                        value   position    error
                        3.6121  A           0.0545
                        3.2381  B           0.0274
                        3.3924  C           0.0176
                    };
                \addplot [blue, ultra thin, pattern=horizontal lines, 
                pattern color=blue,  error bars/.cd, y dir=both, y explicit, error bar style={blue, very thick}] table [x=position, y=value, y error=error] {
                        value   position    error
                        2.6867  A           0.0660
                        2.3649  B           0.0401
                        2.5247  C           0.0184
                    };
                \addplot [green, ultra thin, pattern=crosshatch, 
                pattern color=green, error bars/.cd, y dir=both, y explicit, error bar style={green, very thick}] table [x=position, y=value, y error=error] {
                        value   position    error
                        1.6316  A           0.0556
                        0.9907  B           0.0376
                        0.6634  C           0.0209    
                    };
                \addplot [magenta, ultra thin, pattern= grid, 
                pattern color=magenta, error bars/.cd, y dir=both, y explicit, error bar style={magenta, very thick}] table [x=position, y=value, y error=error] {
                        value   position    error
                        1.5941  A           0.0706
                        0.9494  B           0.0264
                        0.8556  C           0.0254      
                    };
                      \legend{NGUB1, NGUB2, PIA, MDSR, BounceNet}
                    \end{axis}
                  \end{tikzpicture}
		\caption{}
		\label{Fig: vAPUE}
	\end{subfigure}     
	\begin{subfigure}{.49\textwidth}
		\centering
              \begin{tikzpicture}[scale=1]
                \begin{axis}[
                  scale=1.2,
                  ybar=2pt,
                  symbolic x coords={A, B, C},
                  xtick=data,
                  xticklabels={$28$ GHz , $60$ GHz, $73$ GHz},
                  xticklabel style={font={\tiny \boldmath}, rotate=0},
                  minor x tick num=1,
                  minor grid style={dashed},
                  major x tick style={draw=none},
                  minor x tick style={draw=none},
                  enlarge x limits=0.3,
                   ymax=4.3,
                  yticklabel style={font=\tiny \boldmath },
                  ylabel={Throughput per UE (Gbps)},
                  label style={font={\tiny \bfseries}},
                  minor y tick num=1,
                  minor grid style={dashed},
                  enlarge y limits=0.2,
                  major tick length=2pt,
                  minor tick length=1pt,
                  axis line style = thick,
                  legend cell align=left,
                  legend columns=5,
                 legend style={thick, font={\footnotesize}, draw=black, fill=white, at={(0.98,1.12)}},
                 width=\textwidth,
                 bar width=0.09cm, 
                 nodes near coords, 
                 visualization depends on={\thisrow{error} \as \myoffset},
                 every node near coord/.append style={yshift=\myoffset*0.1 cm, black, font={\tiny \boldmath},
                  rotate=90, anchor=west}
                  ]  
              \addplot [black, ultra thin, pattern=north east lines, 
            pattern color=black, error bars/.cd, y dir=both, y explicit, error bar style={black, very thick}] table [x=position, y=value, y error=error] {
                    value   position    error
                    3.4609  A           0.0359
                    3.2402  B           0.0282
                    3.1273  C           0.0243
                }; 
            \addplot [red, ultra thin, pattern=north west lines, 
            pattern color=red, error bars/.cd, y dir=both, y explicit,  error bar style={red, very thick}] table [x=position, y=value, y error=error] {
                    value   position    error
                    3.4581  A           0.0340
                    3.2381  B           0.0274
                    3.1263  C           0.0226
                };
            \addplot [blue, ultra thin, pattern=horizontal lines, 
            pattern color=blue,  error bars/.cd, y dir=both, y explicit, error bar style={blue, very thick}] table [x=position, y=value, y error=error] {
                    value   position    error
                    2.5340  A           0.0437
                    2.3649  B           0.0401
                    2.2920  C           0.0338
                };
            \addplot [green, ultra thin, pattern=crosshatch, 
            pattern color=green, error bars/.cd, y dir=both, y explicit, error bar style={green, very thick}] table [x=position, y=value, y error=error] {
                    value   position    error
                    1.0661  A           0.0355
                    0.9907  B           0.0376
                    0.9687  C           0.0351    
                };
            \addplot [magenta, ultra thin, pattern= grid, 
            pattern color=magenta, error bars/.cd, y dir=both, y explicit, error bar style={magenta, very thick}] table [x=position, y=value, y error=error] {
                    value   position    error
                    0.9656  A           0.0325
                    0.9494  B           0.0264
                    0.9084  C           0.0317     
                };
                \end{axis}
              \end{tikzpicture}
		\caption{}
		\label{Fig: vFE}
	\end{subfigure}
 	\begin{subfigure}{.49\textwidth}
		\centering
              \begin{tikzpicture}[scale=1]
                \begin{axis}[
                  scale=1.2,
                  ybar=2pt,
                  symbolic x coords={A, B, C},
                  xtick=data,
                  xticklabels={SPS=$1$, SPS=$5$, SPS=$10$},
                  xticklabel style={font={\tiny \boldmath}, rotate=0},
                  minor x tick num=1,
                  minor grid style={dashed},
                  major x tick style={draw=none},
                  minor x tick style={draw=none},
                  enlarge x limits=0.3,
                   ymax=4.5,
                  yticklabel style={font=\tiny \boldmath },
                  ylabel={Throughput per UE (Gbps)},
                  label style={font={\tiny \bfseries}},
                  minor y tick num=1,
                  minor grid style={dashed},
                  enlarge y limits=0.2,
                  major tick length=2pt,
                  minor tick length=1pt,
                  axis line style = thick,
                  legend cell align=left,
                  legend columns=5,
                 legend style={thick, font={\footnotesize}, draw=black, fill=white, at={(0.98,1.12)}},
                 width=\textwidth,
                 bar width=0.09cm, 
                 nodes near coords, 
                 visualization depends on={\thisrow{error} \as \myoffset},
                 every node near coord/.append style={yshift=\myoffset*0.1 cm, black, font={\tiny \boldmath},
                  rotate=90, anchor=west}
                  ]  
              \addplot [black, ultra thin, pattern=north east lines, 
            pattern color=black, error bars/.cd, y dir=both, y explicit, error bar style={black, very thick}] table [x=position, y=value, y error=error] {
                    value   position    error
                    3.2402  A           0.0282
                    2.6814  B           0.0457
                    2.6370  C           0.0526
                }; 
            \addplot [red, ultra thin, pattern=north west lines, 
            pattern color=red, error bars/.cd, y dir=both, y explicit,  error bar style={red, very thick}] table [x=position, y=value, y error=error] {
                    value   position    error
                    3.2381  A           0.0274
                    2.7378  B           0.0446
                    2.6283  C           0.0542
                };
            \addplot [blue, ultra thin, pattern=horizontal lines, 
            pattern color=blue,  error bars/.cd, y dir=both, y explicit, error bar style={blue, very thick}] table [x=position, y=value, y error=error] {
                    value   position    error
                    2.3649  A           0.0401
                    2.3762  B           0.0451
                    2.3409  C           0.0571
                };
            \addplot [green, ultra thin, pattern=crosshatch, 
            pattern color=green, error bars/.cd, y dir=both, y explicit, error bar style={green, very thick}] table [x=position, y=value, y error=error] {
                    value   position    error
                    0.9907  A           0.0376
                    0.9626  B           0.0697
                    0.9580  C           0.1290   
                };
            \addplot [magenta, ultra thin, pattern= grid, 
            pattern color=magenta, error bars/.cd, y dir=both, y explicit, error bar style={magenta, very thick}] table [x=position, y=value, y error=error] {
                    value   position    error
                    0.9494  A           0.0264
                    1.0130  B           0.0647
                    1.0504  C           0.0748    
                };
                \end{axis}
              \end{tikzpicture}
		\caption{}
		\label{Fig: vSPS}
	\end{subfigure}
 	\begin{subfigure}{.49\textwidth}
		\centering
              \begin{tikzpicture}[scale=1]
                \begin{axis}[
                  scale=1.2,
                  ybar=2pt,
                  symbolic x coords={A, B, C, D},
                  xtick=data,
                  xticklabels={InH, UMi, UMa, RMa},
                  xticklabel style={font={\tiny \boldmath}, rotate=0},
                  minor x tick num=1,
                  minor grid style={dashed},
                  major x tick style={draw=none},
                  minor x tick style={draw=none},
                  enlarge x limits=0.21,
                   ymax=1.15,
                  yticklabel style={font=\tiny \boldmath },
                  ylabel={Jain's Fairness Index},
                  label style={font={\tiny \bfseries}},
                  minor y tick num=1,
                  minor grid style={dashed},
                  enlarge y limits=0.3,
                  major tick length=2pt,
                  minor tick length=1pt,
                  axis line style = thick,
                  legend cell align=left,
                  legend columns=5,
                 legend style={thick, font={\footnotesize}, draw=black, fill=white, at={(0.98,1.12)}},
                 width=\textwidth,
                 bar width=0.085cm, 
                 nodes near coords, 
                 every node near coord/.append style={black, font={\tiny \boldmath}, /pgf/number format/precision=4, rotate=90, anchor=west}
                  ] 
            
              \addplot [black, ultra thin, pattern=north east lines, 
            pattern color=black] table [x=position, y=value] {
                    value   position   
                    0.9980  A           
                    0.9905  B           
                    0.9906  C           
                    0.9901  D                   
                }; 
            \addplot [red, ultra thin, pattern=north west lines, 
            pattern color=red] table [x=position, y=value] {
                    value   position    
                    0.9976  A           
                    0.9870  B           
                    0.9862  C           
                    0.9874  D           
                };
            \addplot [blue, ultra thin, pattern=horizontal lines, 
            pattern color=blue] table [x=position, y=value] {
                    value   position    
                    0.9941  A           
                    0.9868  B           
                    0.9890  C           
                    0.9903  D           
                };
            \addplot [green, ultra thin, pattern=crosshatch, 
            pattern color=green] table [x=position, y=value] {
                    value   position    
                    0.9444  A           
                    0.6978  B           
                    0.8407  C           
                    0.8326  D           
                };
            \addplot [magenta, ultra thin, pattern= grid, 
            pattern color=magenta] table [x=position, y=value] {
                    value   position    
                    0.9196  A           
                    0.8156  B           
                    0.9229  C           
                    0.8945  D           
                };
                \end{axis}
              \end{tikzpicture}
		\caption{}
		\label{Fig: vJFI}
	\end{subfigure}
        \vspace*{-5mm}
	\caption{The plots show a performance comparison in terms of the per-user average throughput metric (respectively, Jain's Fairness Index (JFI)) in Fig.~\ref{Fig: vAPUE}, Fig.~\ref{Fig: vFE}, Fig.~\ref{Fig: vSPS} (respectively, Fig.~\ref{Fig: vJFI}) under different algorithms for a mmWave network. In Fig.~\ref{Fig: vAPUE},  Fig.~\ref{Fig: vFE}, and  Fig.~\ref{Fig: vSPS} UMi scenario is considered. In Fig.~\ref{Fig: vAPUE}, the parameter used are  $(N_A,N_U)=(4,10), (9,25), (16,40)$,  operating frequency= 60 GHz, number of SPS=1. In Fig.~\ref{Fig: vFE}, the parameters used are $N_A=9$,  $N_U=25$, operating frequency=  28 GHz, 60 GHz, 73 GHz, number of SPS=1. In Fig.~\ref{Fig: vSPS}, the parameters used are $N_A=9$,  $N_U=25$, operating frequency= 60 GHz,  number of SPS=1, 5, 10.  In Fig.~\ref{Fig: vJFI}, the parameters used are $N_A=9$,  $N_U=25$, operating frequency= 60 GHz, number of SPS=10.
}
\label{fig5}
\end{figure}

The result in Fig. \ref{Fig: vScen} shows that our proposed algorithms-- \textit{NGUB1} and \textit{NGUB2}-- perform close to the asymptotically optimal MCMC-based and LIG-based algorithms in terms of per-user average throughput. The results in Figs. \ref{Fig: vScen}-\ref{Fig: vSPS} clearly show that our proposed algorithms-- \textit{NGUB1} and \textit{NGUB2}-- for UE and beam selection outperform the most relevant algorithms-- BounceNet~\cite{jog2019many}, MDSR~\cite{yang2021mdsr} and PIA~\cite{wei2017pose}-- for UE and beam selection in prior work in all mmWave network scenarios in terms of per-user average throughput. The result in Fig. \ref{Fig: vJFI} shows that our algorithms are also fair.

\section{Conclusions and Future Work}
\label{Section: Conclusions}
In this paper, we proved that finding the optimal UE and beam selection in mmWave networks that contain multiple APs and multiple UEs is NP-complete. We proposed two benchmark algorithms--  MCMC-based and LIG-based-- for the problem and proved that the solutions they find are asymptotically optimal. Through simulations, we showed that our proposed NGUB1 and NGUB2 algorithms outperform the most relevant algorithms for UE and beam selection in prior work and that their performance is close to optimal. 

The most relevant algorithms for UE and beam selection in prior work as well as those in this work are designed for the scenario where an AP can communicate with only one UE at any time. Extension of the algorithms designed in this paper to the case where each AP can simultaneously communicate with multiple UEs is a direction for future research.

\appendix
\appendices
\section*{ Proof of Theorem \ref{Theorem: The problem is NP-hard}}
In the following, we describe an NP-complete problem (see Problem \ref{Problem: Finding MIS of a Graph}) using which we prove the NP-hardness of Problem \ref{Problem: Original UE and beam selection Problem}.
Let $G=(\mcV, \mcE)$ be an undirected graph with the set of nodes $\mcV$, the set of edges $\mcE$, and at most three edges at any node. Let $e_{i,j}=1$ (respectively, $e_{i,j}=0$) denote the presence (respectively, absence) of an edge between node $i$ and node $j$. Recall that a set $\widetilde{\mcV} \subseteq \mcV$ is said to be an independent set~\cite{west2001introduction,fleischner2010maximum} if $e_{i,j}=0, \forall i,j \in \widetilde{\mcV}$. Let $\mbV$ denote the set of all independent sets of $G$. Then, $\mcV^* \in \mbV$ is said to be a maximum independent set (MIS)~\cite{west2001introduction,fleischner2010maximum} if $|\mcV^*| \ge |\widetilde{\mcV}|, \forall \widetilde{\mcV} \in \mbV$.

\begin{problem} \label{Problem: Finding MIS of a Graph}
    Find a MIS $\mcV^* \subseteq \mcV$ given $G$.
\end{problem}

 In~\cite{fleischner2010maximum}, it has been shown that Problem \ref{Problem: Finding MIS of a Graph} is NP-complete.

We now show that any instance of Problem \ref{Problem: Finding MIS of a Graph} can be reduced to an instance of Problem \ref{Problem: Original UE and beam selection Problem} in polynomial time as follows. Consider an undirected graph $G=(\mcV, \mcE)$ containing nodes with a maximum degree of three; an arbitrary instance of Problem \ref{Problem: Finding MIS of a Graph} maps to an instance of Problem \ref{Problem: Original UE and beam selection Problem} as follows. A node in the instance of Problem \ref{Problem: Finding MIS of a Graph} maps to an AP-UE-beam triplet in the instance of Problem \ref{Problem: Original UE and beam selection Problem}. We assume that in a mapped AP-UE-beam triplet, the AP corresponds to an AP from the set of APs, from which the UE gets the highest RSS across all beams and the beam corresponds to the best beam between the AP-UE pair.  Further, we assume that the RSS at the UE from the AP in each AP-UE pair is the same and is equal to $nN_0, n \in \mbN$, where $n$ is a fixed positive integer to be specified later and $N_0$ is the noise power. The presence of an edge between two nodes in the instance of Problem \ref{Problem: Finding MIS of a Graph} implies that the AP-UE pairs in the corresponding instance of Problem \ref{Problem: Original UE and beam selection Problem} interfere with each other. Moreover, we assume that if two AP-UE pairs interfere with each other, then the interfering RSS is $\epsilon n N_0$, where $\epsilon \in [0,1]$.

For $\widetilde{\mcV}\subseteq \mcV$, let $N_{\widetilde{\mcV}}^v:=|\{v' \in \widetilde{\mcV}| e_{v,v'}=1, v' \neq v\}|$ denote the number of edges incident on $v \in \widetilde{\mcV}$ whose other endpoint is in $\widetilde{\mcV}$.  Then, assuming $B=1$, the sum rate of the concurrent UEs in the mapped instance of the network in Problem \ref{Problem: Original UE and beam selection Problem} corresponding to an instance $G=(\mcV, \mcE)$ of Problem \ref{Problem: Finding MIS of a Graph} given $\widetilde{\mcV}$, $n$, and $\epsilon$ is given by:
\begin{align}
\mcR^{\ts}(\widetilde{\mcV},n,\epsilon)=\sum_{v \in \widetilde{\mcV}} \log_2 \bigg(1+\frac{n}{1+ N_{\widetilde{\mcV}}^v \epsilon n} \bigg). \label{Eq: Normalized sum rate given network}
\end{align}

We define two functions and establish some results, which will be useful in establishing subsequent results of this section. Let $f(m,n,\epsilon)=\frac{m \epsilon n+1}{(m \epsilon+1) n+1}\cdot \frac{((m-1)  \epsilon+1) n+1}{(m-1) \epsilon n+1}$ and $g(m,n,\epsilon)=\frac{m \epsilon n+1}{(m \epsilon+1) n+1}$ with $m \in \{ 1,2,3\}, n \in \mathbb{N}, \epsilon \in (0,1)$.
\begin{lemma} \label{Lemma: Bound on inside term}
    The following results hold: 
    \begin{itemize}
        \item[(i)] $\lim_{n \uparrow \infty, \epsilon \uparrow 1} g(1,n,\epsilon)  f(1,n,\epsilon) =\infty$.
        \item[(ii)] $\lim_{n \uparrow \infty, \epsilon \uparrow 1 } g(2,n,\epsilon) f(m_1,n,\epsilon)  f(m_2,n,\epsilon)\\ \ge \frac{32}{27}, m_1, m_2 \in \{ 1,2\}$.
        \item[(iii)]  $\lim_{n \uparrow \infty, \epsilon \uparrow 1}  g(3,n,\epsilon)f(m_1,n,\epsilon)  f(m_2,n,\epsilon) f(m_3,n,\epsilon)\\ \ge \frac{2187}{2048},  m_1, m_2, m_3 \in \{ 1,2,3\}$.
    \end{itemize}
\end{lemma} 
\begin{IEEEproof} 
    It can be seen that as $n \uparrow \infty, \text{ and } \epsilon \uparrow 1$, $f(1,n,\epsilon)\uparrow \infty$, $ f(2,n,\epsilon) \uparrow \frac{4}{3}$, $f(3,n,\epsilon) \uparrow \frac{9}{8}$, $g(1,n,\epsilon) \uparrow \frac{1}{2}$, $g(2,n,\epsilon) \uparrow \frac{2}{3}$, and $g(3,n,\epsilon) \uparrow \frac{3}{4}$. The result follows from these facts.
\end{IEEEproof}

In the following, addition (respectively, deletion) of a node to (respectively, from) a subset of nodes in an instance of Problem \ref{Problem: Finding MIS of a Graph} corresponds to addition (respectively, deletion) of an AP-UE pair in the corresponding instance of Problem \ref{Problem: Original UE and beam selection Problem}. An independent subset of nodes in an instance of Problem \ref{Problem: Finding MIS of a Graph} corresponds to a non-interfering subset of AP-UE pairs in the corresponding instance of Problem \ref{Problem: Original UE and beam selection Problem}. Note that given an instance $G=(\mcV,\mcE)$ of Problem \ref{Problem: Finding MIS of a Graph} and $\widetilde{\mcV} \subseteq \mcV$, we can find the corresponding sum rate using (\ref{Eq: Normalized sum rate given network}) by obtaining the corresponding instance of Problem \ref{Problem: Original UE and beam selection Problem}.

Let $\widetilde{\mbV}:=\{ {\mcV}' \subseteq \mcV| \sum_{v \in\mcV'} N_{\mcV'}^v>0 \}$ be the collection of all non-independent subsets of $\mcV$.
\begin{lemma} \label{Lemma: Dropping leads to INSR}
    In every $\widetilde{\mcV} \in \widetilde{\mbV}$, there must exist a node such that dropping it leads to an increment in the sum rate. 
\end{lemma}
\begin{IEEEproof} 
    Given any $\widetilde{\mcV} \in \mcV$, let $v \in \widetilde{\mcV}$ be such that $v \in \argmax_{v'} N_{\widetilde{\mcV}}^{v'}$. Recall that any node has at most three edges. Thus, there are three possible cases-- $N_{\widetilde{\mcV}}^v=1$, $N_{\widetilde{\mcV}}^v=2$, and $N_{\widetilde{\mcV}}^v=3$. We show that in all these cases, there exist $n \in \mathbb{N}$ and $\epsilon \in (0,1)$ such that dropping $v$ leads to an increment in the sum rate.  

    Let $I_m^{\tde}$ denote the increment in the rate of the UE corresponding to a node having $m \in \{1,2,3 \}$ edges when an edge of the node is dropped. Note that $I_m^{\tde}=\log_2(f(m,n,\epsilon))$. Let $D_m^{\tdn}$ denote the decrement in the rate of the UE corresponding to a node when the node gets dropped and the node has $m$ edges. Note that $D_m^{\tdn}=\log_2(1/g(m,n,\epsilon))$. Let $m_{\td} = N_{\widetilde{\mcV}}^v$ be the number of edges at the dropped node $v$. 
    
    \textbf{Case 1: $m_{\td}=1$}\\
    Let $v'$ be the node that has an edge with $v$. Let $m_1$ be the number of edges at node $v'$. Then dropping node $v$ from $\widetilde{\mcV}$ leads to the following change in the sum rate when $n \uparrow \infty$ and $\epsilon \uparrow 1$:
    \begin{align*}
        &\lim_{n \uparrow \infty, \epsilon \uparrow 1 } \Big(\mcR^{\ts}(\{ \widetilde{\mcV} \setminus v \},n,\epsilon)-\mcR^{\ts}(\widetilde{\mcV},n,\epsilon)\Big)\\
        =& \lim_{n \uparrow \infty, \epsilon \uparrow 1 } (I_{m_1}^{\tde}-D_{m_{\td}}^{\tdn})\\
        =& \lim_{n \uparrow \infty, \epsilon \uparrow 1 } \log_2(f(m_1,n,\epsilon)g(m_{\td},n,\epsilon))\\
        =& \infty.
    \end{align*}
   The first equality holds since the change in the sum rate happens only due to the changes in the rates of the UEs corresponding to nodes $v$ and $v'$. The second equality follows from the definitions of $I_m^{\tde}$ and $D_m^{\tdn}$.  The third equality follows from Lemma \ref{Lemma: Bound on inside term} and the fact that $m_{\td}=1$ and $m_1=1$.

        \textbf{Case 2: $m_{\td}=2$}\\
    Let $v'$ and $v''$ be the nodes that have an edge with $v$. Let $m_1$ and $m_2$ be the numbers of edges at nodes $v'$ and $v''$, respectively. Then dropping node $v$ from $\widetilde{\mcV}$ leads to the following change in the sum rate when $n \uparrow \infty$ and $\epsilon \uparrow 1$: 
    \begin{align*}
        &\lim_{n \uparrow \infty, \epsilon \uparrow 1 } \Big(\mcR^{\ts}(\{ \widetilde{\mcV} \setminus v \},n,\epsilon)-\mcR^{\ts}(\widetilde{\mcV},n,\epsilon)\Big)\\
        =& \lim_{n \uparrow \infty, \epsilon \uparrow 1 } (I_{m_1}^{\tde}+I_{m_2}^{\tde}-D_{m_{\td}}^{\tdn})\\
        =& \lim_{n \uparrow \infty, \epsilon \uparrow 1 } \log_2(f(m_1,n,\epsilon)f(m_2,n,\epsilon)g(m_{\td},n,\epsilon))\\
        \ge& \log_2(32/27).
    \end{align*}
   The first equality holds since the change in the sum rate happens only due to the changes in the rates of the UEs corresponding to nodes $v$, $v'$, and $v''$. The second equality follows from the definitions of $I_m^{\tde}$ and $D_m^{\tdn}$.  The third equality follows from Lemma \ref{Lemma: Bound on inside term} and the fact that $m_{\td}=2$ and $m_1,m_2 \in \{1,2\}$.

            \textbf{Case 3: $m_{\td}=3$}\\
     Let $v'$, $v''$, and $v'''$ be the nodes that have an edge with $v$. Let $m_1$, $m_2$, and $m_3$ be the numbers of edges at nodes $v'$, $v''$, and  $v'''$, respectively. Then dropping node $v$ from $\widetilde{\mcV}$ leads to the following change in the sum rate when $n \uparrow \infty$ and $\epsilon \uparrow 1$: 
    \begin{align*}
        &\lim_{n \uparrow \infty, \epsilon \uparrow 1 } \Big(\mcR^{\ts}(\{ \widetilde{\mcV} \setminus v \},n,\epsilon)-\mcR^{\ts}(\widetilde{\mcV},n,\epsilon)\Big)\\
        =& \lim_{n \uparrow \infty, \epsilon \uparrow 1 } (I_{m_1}^{\tde}+I_{m_2}^{\tde}+I_{m_3}^{\tde}-D_{m_{\td}}^{\tdn})\\
        =& \lim_{n \uparrow \infty, \epsilon \uparrow 1 } \log_2(f(m_1,n,\epsilon)f(m_2,n,\epsilon)f(m_3,n,\epsilon)g(m_{\td},n,\epsilon))\\
        \ge & \log_2(2187/2048).
    \end{align*}
   The first equality holds since the change in the sum rate happens only due to the changes in the rates of the UEs corresponding to nodes $v$, $v'$, and $v''$. The second equality follows from the definitions of $I_m^{\tde}$ and $D_m^{\tdn}$.  The third equality follows from Lemma \ref{Lemma: Bound on inside term} and the fact that $m_{\td}=3$ and $m_1,m_2, m_3 \in \{1,2,3\}$.   
\end{IEEEproof}

Lemma \ref{Lemma: Dropping leads to INSR} says that in Problem \ref{Problem: Finding MIS of a Graph}, for every subset of the set of nodes of a graph that is not an independent set, there exists at least one node, such that dropping it increases the sum rate of concurrent transmissions of the corresponding AP-UE pairs in Problem \ref{Problem: Original UE and beam selection Problem}.  

\begin{lemma} \label{Lemma: MSR implies MIS}
    If the set of AP-UE pairs in Problem \ref{Problem: Original UE and beam selection Problem} corresponding to the set $\mcV^* \subseteq \mcV$ in Problem \ref{Problem: Finding MIS of a Graph} maximizes the  sum rate, then $\mcV^*$ must be a MIS.
\end{lemma}
\begin{IEEEproof} 
  To reach a contradiction, suppose the set of AP-UE pairs in Problem \ref{Problem: Original UE and beam selection Problem} corresponding to the set $\mcV^*$ in Problem \ref{Problem: Finding MIS of a Graph} maximizes the sum rate, but $\mcV^*$ is not a MIS. Then one of the following two cases must be true:  (i) first, $\mcV^*$ is an independent set, but not of maximum cardinality, (ii) second, $\mcV^*$ is not an independent set. 
  
  First, consider the case where $\mcV^*$ is an independent set, but not of maximum cardinality. In this case, we can replace $\mcV^*$ with an independent set of maximum cardinality; this leads to at least $\log_2(1+n)$ increment in the sum rate. This contradicts the fact that the set of AP-UE pairs in Problem \ref{Problem: Original UE and beam selection Problem} corresponding to the set $\mcV^*$ in Problem \ref{Problem: Finding MIS of a Graph} maximizes the sum rate. 
  
  Now consider the case where $\mcV^*$ is not an independent set. Then  by Lemma \ref{Lemma: Dropping leads to INSR}, there exists a node $v^* \in \mcV^*$  such that dropping it from $\mcV^*$ increases the sum rate. 
  This contradicts the fact that the set of AP-UE pairs in Problem \ref{Problem: Original UE and beam selection Problem} corresponding to the set $\mcV^*$ in Problem \ref{Problem: Finding MIS of a Graph} maximizes the sum rate. Hence, $\mcV^*$ must be a MIS.
\end{IEEEproof}
Lemma \ref{Lemma: MSR implies MIS} says that the subset of nodes in an instance of Problem \ref{Problem: Finding MIS of a Graph} corresponding to a concurrent set of AP-UE pairs that maximizes the sum rate in the corresponding instance of Problem \ref{Problem: Original UE and beam selection Problem} must be a MIS.

\begin{lemma} \label{Lemma: MIS implies MSR}
    If $\mcV^* \subseteq \mcV$ is a MIS, then the set of AP-UE pairs in Problem \ref{Problem: Original UE and beam selection Problem} corresponding to the set $\mcV^*$ in Problem \ref{Problem: Finding MIS of a Graph} maximizes the sum rate.
\end{lemma}
\begin{IEEEproof} 
     The sum rate of the AP-UE pairs in Problem \ref{Problem: Original UE and beam selection Problem} corresponding to an independent set of cardinality $k'$  in Problem \ref{Problem: Finding MIS of a Graph} is $k' \log_2(1+n)$. Hence, the set of AP-UE pairs in Problem \ref{Problem: Original UE and beam selection Problem} corresponding to a MIS in Problem \ref{Problem: Finding MIS of a Graph} maximizes the sum rate over all the sets of AP-UE pairs in Problem \ref{Problem: Original UE and beam selection Problem} that correspond to independent sets of $\mcV$ in Problem \ref{Problem: Finding MIS of a Graph}. Also, from Lemma \ref{Lemma: Dropping leads to INSR}, it follows that starting from any non-independent subset of $\mcV$, and in each step, by deleting a node with the maximum number of edges incident on it, we can reduce the subset to an independent set in a finite number of steps and each step increases the sum rate. It follows that the set of AP-UE pairs in Problem \ref{Problem: Original UE and beam selection Problem} corresponding to a MIS in Problem \ref{Problem: Finding MIS of a Graph} maximizes the sum rate over all possible sets of AP-UE pairs in Problem \ref{Problem: Original UE and beam selection Problem}.     
\end{IEEEproof}
Lemma \ref{Lemma: MIS implies MSR} says that if a subset of nodes in an instance of Problem \ref{Problem: Finding MIS of a Graph} is a MIS, then it corresponds to a set of AP-UE pairs in Problem \ref{Problem: Original UE and beam selection Problem} that maximizes the sum rate.
\begin{IEEEproof}[Proof of Theorem \ref{Theorem: The problem is NP-hard}]
 The decision version of Problem \ref{Problem: Original UE and beam selection Problem} is: given a number $R_0$, does there exist a set of beam-UE-AP triplets $\mcF$ given $\mcS, \bw$ of a schedule such that $R_{\mcS,\bw, \mcF} \ge R_0?$ Given $\mcF, \mcS, \bw$ of a schedule, we can check whether  $R_{\mcS,\bw, \mcF} \ge R_0$ holds or not in polynomial time; so Problem \ref{Problem: Original UE and beam selection Problem} lies in class NP~\cite{kleinberg2006algorithm}.  
 
 To show that Problem \ref{Problem: Original UE and beam selection Problem} is NP-complete, it remains to show that Problem \ref{Problem: Original UE and beam selection Problem} is NP-hard. Now, from Lemmas \ref{Lemma: MSR implies MIS} and \ref{Lemma: MIS implies MSR} and the fact that Problem \ref{Problem: Finding MIS of a Graph} has been shown to be NP-complete in~\cite{fleischner2010maximum}, it follows that Problem \ref{Problem: Original UE and beam selection Problem} is NP-hard. The result follows. 
\end{IEEEproof}
\section*{Proof of Theorem \ref{Theorem: MCMC-based Algo asymptotically achieve optimum UE and beam selection}}

 \begin{IEEEproof}[Proof of Lemma \ref{Lemma: MCMC unique stationary distribution}]
    Note that Algorithm \ref{Algorithm: MCMC} induces a Markov chain $\{\bb^i\}_{i\in \mbN}$ over the space of beam vectors $\mbB$, where $\bb^i$ is the beam vector selected by Algorithm \ref{Algorithm: MCMC} in the $i$th iteration. 
    Let $p_{\Hat{\bb}, \Tilde{\bb}}$ be the transition probability $\Pr(\bb^i=\Tilde{\bb}|\bb^{i-1}=\Hat{\bb}), \forall i \in \mbN$ from state $\Hat{\bb}$ to state $\Tilde{\bb}$. Let $\mbB_{\Hat{\bb}}:=\{ \bb \in \mbB| \sum_{a \in \mcA} \indicator_{\{{\bb}_a \neq \Hat{\bb}_a \}} \le 1 \}$, where $\indicator_{\{ \cdot \}}$ is an indicator function. Note that  Algorithm \ref{Algorithm: MCMC} chooses a beam vector $\Tilde{\bb}$ from $\mbB_{\Hat{\bb}}$ with probability $\frac{1}{N_B N_A}$ and accepts it as the current beam vector for the $i$th iteration with probability one (respectively, $e^{\alpha (R_{\Tilde{\bb}} - R_{\Hat{\bb}})}$) if $R_{\Tilde{\bb}} > R_{\Hat{\bb}}$ (respectively, $R_{\Tilde{\bb}} \le R_{\Hat{\bb}}$). Hence: $p_{\Hat{\bb},\Tilde{\bb}}=$
    \begin{align*}
         & \begin{cases}
            0, & \Tilde{\bb} \notin \mbB_{\Hat{\bb}},\\
            \frac{1}{N_B N_A}, & \Tilde{\bb} \in \mbB_{\Hat{\bb}} \text{ and }  R_{\Tilde{\bb}} > R_{\Hat{\bb}}, \\
            \frac{e^{\alpha (R_{\Tilde{\bb}} - R_{\Hat{\bb}})}}{N_B N_A},  & \Tilde{\bb} \in \{ \mbB_{\Hat{\bb}} \setminus \Hat{\bb} \}\text{ and }  R_{\Tilde{\bb}} \le R_{\Hat{\bb}}, \\
           \frac{1}{N_B N_A}+ \sum\limits_{\Tilde{\bb} \in \{ \mbB_{\Hat{\bb}} \setminus \Hat{\bb} \}| R_{\Tilde{\bb}} \le R_{\Hat{\bb}}}\\ \frac{1-e^{\alpha (R_{\Tilde{\bb}} - R_{\Hat{\bb}})}}{N_B N_A}, & \Tilde{\bb}=\Hat{\bb}.  
        \end{cases}
    \end{align*}

    To prove that the algorithm achieves a unique stationary distribution, we have to show that the Markov chain induced by the algorithm is aperiodic and irreducible. Note from the expression of transition probability that the chain remains in the current state with a positive probability; hence, the chain is aperiodic. The algorithm can visit any state from any state in a finite number of steps, and each transition happens with a positive probability, and therefore the chain is irreducible. Now, we prove that $\pi(\cdot)$ is the unique stationary distribution of the chain $\{\bb^i \}_{i \in \mbN}$ by showing that it satisfies the detailed balance equations $\pi (\Tilde{\bb}) p_{\Tilde{\bb},\Hat{\bb}}=\pi (\Hat{\bb}) p_{\Hat{\bb},\Tilde{\bb}}, \forall \Tilde{\bb},\Hat{\bb} \in \mbB$. 
    Given any $\Tilde{\bb} \in \mbB$, one of the following four cases must be true. We show that for each of these cases, the detailed balance equations are satisfied.

    \textbf{Case 1:} $\Hat{b} \notin \mbB_{\Tilde{\bb}}$\\
     It is easy to see that $\Hat{b} \notin \mbB_{\Tilde{\bb}}$ implies $\Tilde{b} \notin \mbB_{\Hat{\bb}}$ and therefore both $p_{\Hat{\bb},\Tilde{\bb}}$ and $p_{\Tilde{\bb},\Hat{\bb}}$ are zero. Thus, both sides of the detailed balance equation are zero.
    
    \textbf{Case 2:} $\Hat{b}=\Tilde{\bb}$\\
    It is clear that in this case, both sides of the detailed balance equation are the same.
    
    \textbf{Case 3:} $\Hat{b} \in \{\mbB_{\Tilde{\bb}} \setminus \Tilde{\bb} \} \text{ and } R_{\Tilde{\bb}} > R_{\Hat{\bb}}$\\
    It is easy to see that $\Hat{b} \in \{\mbB_{\Tilde{\bb}} \setminus \Tilde{\bb} \}$ implies $\Tilde{b} \in \{\mbB_{\Hat{\bb}} \setminus \Hat{\bb} \}$. So in this case, $p_{\Hat{\bb},\Tilde{\bb}}=\frac{1}{N_B N_A}$ and $p_{\Tilde{\bb},\Hat{\bb}}=\frac{e^{\alpha (R_{\Hat{\bb}}-R_{\Tilde{\bb}})}}{N_B N_A}$. Therefore, both sides of the detailed balance equation are equal and are equal to $\frac{e^{\alpha R_{\Hat{\bb}}}}{N_B N_A \sum_{\bb \in \mbB} e^{\alpha R_{\bb}} }$.
    
    \textbf{Case 4:} $\Hat{b} \in \{\mbB_{\Tilde{\bb}} \setminus \Tilde{\bb} \} \text{ and } R_{\Tilde{\bb}} \le R_{\Hat{\bb}}$\\
    Similar to Case 3, it can be shown that in this case, both sides of the detailed balance equation are equal and are equal to $\frac{e^{\alpha R_{\Tilde{\bb}}}}{N_B N_A \sum_{\bb \in \mbB} e^{\alpha R_{\bb}} }$.
\end{IEEEproof}

\begin{IEEEproof}[Proof of Lemma \ref{Lemma: MCMC iterates belongs to set of optimal beam vectors}]
    Lemma \ref{Lemma: MCMC unique stationary distribution} states that Algorithm \ref{Algorithm: MCMC} achieves the unique stationary distribution $\pi (\bb)=\frac{e^{\alpha R_{\bb}}}{\sum_{\bb' \in \mbB} e^{\alpha R_{\bb'}}}, \, \forall \bb \in \mbB$ over the space of all possible beam vectors.
    For sufficiently large $\alpha$, $e^{\alpha R_{\bb^{\topt}}} \gg e^{\alpha R_{\bb}}, \, \forall \bb^{\topt} \in \mbB^{\topt},  \bb \in \mbB \setminus \mbB^{\topt}$.
    Therefore:
    \begin{align*}
    \lim_{\alpha \uparrow \infty} \pi(z)=
        \begin{cases}
           \frac{1}{|\mbB^{\topt}|}, & \forall \bb  \in \mbB^{\topt}, \\
           0, & \forall \bb \in \mbB \setminus \mbB^{\topt}.
        \end{cases}
    \end{align*}  
\end{IEEEproof}
\begin{IEEEproof}[Proof of Theorem \ref{Theorem: MCMC-based Algo asymptotically achieve optimum UE and beam selection}]
    Recall that $\argmax_{(\bb,\bu) \in \{ \mbB \times \mbU \}} R_{\mcS,\bw,\bb,\bu}=\argmax_{\bb \in \mbB}\Big(\argmax_{\bu \in \mbU} R_{\mcS,\bw,\bb,\bu} \Big)$, i.e., maximization of the weighted sum rate can be done in two steps, first, maximization over all possible UE vectors for each beam vector, and then maximization over all possible beam vectors.  
    Note that since $R_{\bb}=\max_{\bu \in \mbU} R_{\mcS,\bw,\bb,\bu}$, in step 6 of Algorithm \ref{Algorithm: MCMC}, we find the optimal UE vector for the candidate beam vector $\bb^i$ of each iteration $i$ in polynomial time. Thus, it remains to show that Algorithm \ref{Algorithm: MCMC} finds the optimal beam vector. Lemma \ref{Lemma: MCMC iterates belongs to set of optimal beam vectors} states that for a sufficiently large value of $\alpha$, the stationary distribution has positive probabilities only for beam vectors belonging to the set of optimal beam vectors $\mbB^{\topt}$, which implies that as $i \uparrow \infty$, $\bb^i \in \mbB^{\topt}$. The result follows.
\end{IEEEproof}
\section*{Proof of Theorem \ref{Theorem: G is CPG}}
\begin{IEEEproof}[Proof of Theorem \ref{Theorem: G is CPG}]
To prove that $\mathbf{G}$ is a potential game with potential function $\Psi$, we need to show that $\Psi$ satisfies the condition in Definition \ref{Definition: Constrained Potential Game}.   The utility of player $l$ depends only on its own strategy and the strategy of its neighbors belonging to the set $\mcN_l^{\tut}$; hence, $V_l(z_l,z_{-l})=V_l(z_l,z_{\mcN_l^{\tut}})$ and $\Psi(z_l, z_{-l})=\sum_{l \in \mcL} V_l(z_l,z_{\mcN_l^{\tut}})$. Now, 
    \begin{align*}
        &\Psi({z'}_l, z_{-l})-\Psi(z_l, z_{-l})\\
        =&V_l({z'}_l,z_{\mcN_l^{\tut} })-V_l(z_l,z_{\mcN_l^{\tut} })\\
        &+\sum_{l' \in \{\mcN_l^{\ta} \bigcup \mcN_l^{\tu} \bigcup \mcN_l^{\tc}\}} V_{l'}({z}_{l'},{z'}_{\mcN_{l'}^{\tut} })-V_{l'}(z_{l'},z_{\mcN_{l'}^{\tut} })\\
        &+\sum_{l' \in \mcL \setminus \{l \bigcup\mcN_l^{\ta} \bigcup \mcN_l^{\tu} \bigcup\mcN_l^{\tc} \} } V_{l'}({z}_{l'},{z'}_{\mcN_{l'}^{\tut} })-V_{l'}(z_{l'},z_{\mcN_{l'}^{\tut} })\\
        =&V_l({z'}_l,z_{\mcN_l^{\tut} })-V_l(z_l,z_{\mcN_l^{\tut} })\\
        &+\sum_{l' \in \{\mcN_l^{\ta} \bigcup \mcN_l^{\tu} \bigcup \mcN_l^{\tc}\}} V_{l'}({z}_{l'},{z'}_{\mcN_{l'}^{\tut} })-V_{l'}(z_{l'},z_{\mcN_{l'}^{\tut} })\\
        = &Y_l({z'}_l, z_{\mcN_l^{\tp}})-Y_l({z}_l, z_{\mcN_l^{\tp}})\\
        =&Y_l({z'}_l, z_{-l})-Y_l(z_l, z_{-l})
    \end{align*}
    The second equality holds since the utilities of players $l' \in \mcL \setminus \{l \bigcup \mcN_l^{\ta} \bigcup \mcN_l^{\tu} \bigcup\mcN_l^{\tc} \} $ do not depend on the strategy of $l$. The third equality follows by the definition of the payoff function of player $l$. The fourth equality follows from the fact that the payoff function of player $l$ depends on the strategies of the players belonging to set $\{l \bigcup \mcN_l^{\tp}\}$.

   Let $z^i:=\{z_l^i, \forall l \in \mcL\}$ be the strategy profile chosen by the players in step $i$. Let $z^0=\{z^0_l=0, \forall l \in \mcL\}$. In each step $i \in \{1, 2, \cdots\}$, $z^i \in \mcZ_{\mcL}$ is chosen such that it satisfies two conditions: (i) $z^i$ and $z^{i-1}$ differ exactly for one player, i.e., exactly one player changes its strategy in step $i$, and (ii) the player, say $l$, which changes its strategy, improves its payoff, i.e., $Y_l(z^i)>Y_l(z^{i-1})$. Since $\mathbf{G}$ is a potential game, $Y_l(z^i)>Y_l(z^{i-1}) \implies \Psi(z^i)>\Psi(z^{i-1})$. Thus, each step improves the potential function value. Since  $\mcZ_{\mcL}$ has finite cardinality, the improvement of the potential function happens until some finite step, say $t$.  Thus, after the $t$th step, $\Psi(z^t)$ cannot be improved by changing the strategy of exactly one player, i.e.,  for all $l \in \mcL$ and for all ${z'}_l \in \mcZ_l$, we have $Y_l(z^t)>Y_l({z'}_l,z^t_{-l})$. Hence, $z^t$ is a pure strategy NE for the game $\mathbf{G}$.

   Let $z^{\ast} \in \mcZ_{\mcL}$ be the global maximum of $\Psi$. Then, for all $l \in \mcL$ and for all $z_l \in \mcZ_l$, $\Psi({z^{\ast}}_l,{z^{\ast}}_{-l}) \ge \Psi(z_l,{z^{\ast}}_{-l})$. Since $\mathbf{G}$ is a potential game with potential function $\Psi$, for all $l \in \mcL$ and for all $z_l \in \mcZ_l$, $\Psi({z^{\ast}}_l,{z^{\ast}}_{-l}) \ge \Psi(z_l,{z^{\ast}}_{-l})$ implies $Y_l({z^{\ast}}_l,{z^{\ast}}_{-l}) \ge Y_l(z_l,{z^{\ast}}_{-l})$. Thus, $z^*$ is a pure strategy NE of game $\mathbf{G}$.
\end{IEEEproof}

\section*{Proof of Theorem \ref{Theorem: LIG attain global opt. of PSI}}
\begin{IEEEproof}[Proof of Lemma \ref{Lemma: stationary distribution LIG}]
     Let $\{Z(i) \in \mcZ_{\mcL}\}_{i\ge 1}$ (respectively, $Z(0) \in \mcZ_{\mcL}$) denote the strategy profile of the players after the $i$th update by Algorithm \ref{Algorithm: LIG based UE and Beam Selection} (respectively, the initial strategy profile  of the players). Note that $\{Z(i)\}_{i\ge 0}$ is a Markov chain. We first prove that the Markov chain $\{Z(i)\}_{i \ge 0}$ has a unique stationary distribution over the strategy space $\mcZ_{\mcL}$ and then we prove that $\pi(\cdot)$ is the unique distribution.

    To prove that $\{Z(i)\}_{i \ge 0}$ has a unique stationary distribution over the strategy space, we have to prove that it is aperiodic and irreducible. Recall that Algorithm \ref{Algorithm: LIG based UE and Beam Selection} starts with some initial strategy profile of the players and in each iteration $i \ge 1$, updates the strategies of an independent set of players $\Hat{\mcL}(i) \in \mbL$. Each player $l \in \Hat{\mcL}(i)$ follows the Gibbs distribution $\mu_l(i)$ to get its updated strategy. Note that with a positive probability, a player does not change its strategy;  hence, with a positive probability, none of the players change their strategies. Hence, the Markov chain $\{Z(i)\}_{i \ge 0}$ is aperiodic. Let $\Hat{z}, \Tilde{z} \in \mcZ_{\mcL}$ be any two strategy profiles of the players in the strategy space. It is possible to go to $\Tilde{z}$ (respectively, $\Hat{z}$) from $\Hat{z}$ (respectively, $\Tilde{z}$) in a finite number of steps since the strategy space is finite. Since in each step, an independent set of players update their strategies, it is easy to see that Algorithm \ref{Algorithm: LIG based UE and Beam Selection} follows each of these steps with some positive probability and therefore the chain $\{Z(i)\}_{i \ge 0}$ can visit any strategy profile from any strategy profile in the strategy space with a positive probability. Thus, $\{Z(i)\}_{i \ge 0}$ is irreducible.
    
    Now, we prove that $\pi(\cdot)$ is the unique stationary distribution of the chain $\{Z(i)\}_{i \ge 0}$ by proving that it satisfies the detailed balance equations. Let $P_{\Hat{z},\Tilde{z}}:=\Pr(Z(i)=\Tilde{z}|Z(i-1)=\Hat{z}), \, \forall i \ge 1, \forall \Hat{z},\Tilde{z} \in \mcZ_{\mcL}$ denote the transition probability from state $\Hat{z}$ to state $\Tilde{z}$. Then $\pi(\cdot)$ of the chain $\{Z(i)\}_{i \ge 0}$ satisfies the detailed balance equations if $\pi(\Hat{z})P_{\Hat{z},\Tilde{z}}=\pi(\Tilde{z})P_{\Tilde{z},\Hat{z}}, \, \forall \Tilde{z},\Hat{z} \in \mcZ_{\mcL}$. Let $\Hat{z}_l$ (respectively, $\Tilde{z}_l$) denote the $l$th component of $\Hat{z}$ (respectively, $\Tilde{z}$) and correspond to the strategy of player $l$ before (respectively, after) the $i$th strategy update. Let: 
    \begin{align*}
        P_{\Hat{z}_l, \Tilde{z}_l}:=\begin{cases}
            \frac{e^{\beta Y_l\big(\Tilde{z}_l,\Hat{z}_{\mcN_l^{\tp}}\big)}}{\sum_{z_l \in \mcZ_l} e^{\beta Y_l\big({z}_l,\Hat{z}_{\mcN_l^{\tp}}\big)} }, & l \in \Hat{\mcL}(i),\\
            1, & \Hat{z}_l=\Tilde{z}_l, l \in \mcL \setminus \Hat{\mcL}(i),\\
            0, & \Hat{z}_l \neq \Tilde{z}_l, l \in \mcL \setminus \Hat{\mcL}(i).
        \end{cases}
    \end{align*}
    Then, $P_{\Hat{z},\Tilde{z}}=\frac{1}{|\mbL|}\prod_{l \in \Hat{\mcL}(i)} P_{\Hat{z}_l,\Tilde{z}_l}$.  Note that the probability with which a mutually independent set of players $\Hat{\mcL}(i) \in \mbL$ gets selected by Algorithm \ref{Algorithm: LIG based UE and Beam Selection} in any iteration is $\frac{1}{|\mbL|}$. Now, we can write a proof of the detailed balance equation from its LHS to its RHS as follows.
    \begin{align} \label{Eq: Detailed balance equation proof}
        \pi(\Hat{z}) P_{\Hat{z},\Tilde{z}}&=\alpha_1 e^{\beta \Psi(\Hat{z})} \prod_{l \in \Hat{\mcL}(i)} e^{\beta Y_l\big(\Tilde{z}_l,\Hat{z}_{\mcN_l^{\tp}}\big)} \nonumber \\
        &=\alpha_1  e^{\beta \big(\Psi(\Hat{z})+ \sum_{l \in \Hat{\mcL}(i)} Y_l\big(\Tilde{z}_l,\Hat{z}_{\mcN_l^{\tp}}\big)\big)} \nonumber \\
        &=\alpha_2  e^{\beta \big(\Psi(\Hat{z})+ \sum_{l \in \Hat{\mcL}(i)} Y_l\big(\Tilde{z}_l,\Hat{z}_{\mcN_l^{\tp}}\big)\big)} \\
        &=\alpha_2 e^{\beta \big(\Psi(\Tilde{z})+ \sum_{l \in \Hat{\mcL}(i)} Y_l\big(\Hat{z}_l,\Tilde{z}_{\mcN_l^{\tp}}\big)\big)} \nonumber \\
        &=\alpha_2 e^{\beta \Psi(\Tilde{z})} \prod_{l \in \Hat{\mcL}(i)} e^{\beta Y_l\big(\Hat{z}_l,\Tilde{z}_{\mcN_l^{\tp}}\big)} \nonumber \\
        &=\pi(\Tilde{z}) P_{\Tilde{z},\Hat{z}} \nonumber
    \end{align}
    In (\ref{Eq: Detailed balance equation proof}), $\alpha_1=\frac{1}{|\mbL|} \frac{1}{\sum_{z \in \mcZ_{\mcL}} e^{\beta \Psi(z)}}  \prod_{l \in \Hat{\mcL}(i)} \frac{1}{\sum{z_l \in \mcZ_l} e^{\beta Y_l\big(z_l, \Hat{z}_{\mcN_l^{\tp}}\big)}}$ and $\alpha_2=\frac{1}{|\mbL|} \frac{1}{\sum_{z \in \mcZ_{\mcL}} e^{\beta \Psi(z)}}  \prod_{l \in \Hat{\mcL}(i)} \frac{1}{\sum{z_l \in \mcZ_l} e^{\beta Y_l\big(z_l, \Tilde{z}_{\mcN_l^{\tp}}\big)}}$. 
It is easy to see that the first, the second, the fifth, and the sixth equality holds. Note that the strategies of players $l'\in \mcN_l^{\tp}$ remain the same when Algorithm \ref{Algorithm: LIG based UE and Beam Selection} updates the strategy of player $l \in \Hat{\mcL}(i)$, and therefore $\Hat{z}_{\mcN_l^{\tp}}=\Tilde{z}_{\mcN_l^{\tp}}$ and $\alpha_1=\alpha_2$. Thus, the third equality in (\ref{Eq: Detailed balance equation proof}) holds. To show that the fourth equality in (\ref{Eq: Detailed balance equation proof}) holds, we have to prove that
$\Psi(\Hat{z})+\sum_{l \in \Hat{\mcL}(i)} Y_l(\Tilde{z}_l, \Tilde{z}_{\mcN_l^{\tp}})=\Psi(\Tilde{z})+\sum_{l \in \Hat{\mcL}(i)} Y_l(\Hat{z}_l, \Hat{z}_{\mcN_l^{\tp}})$. Note that the change in the potential function obtained when all players $l \in \Hat{\mcL}(i)$ update their strategies simultaneously as in Algorithm \ref{Algorithm: LIG based UE and Beam Selection} is the same as the sum of the  changes in the potential function obtained when each player $l \in \Hat{\mcL}(i)$ updates its strategy one by one. Consider a path $z^0 \rightarrow z^1 \rightarrow \cdots \rightarrow z^{|\Hat{\mcL}(i)|}$  with $z^0=\Hat{z}$ and $z^{|\Hat{\mcL}(i)|}=\Tilde{z}$, where the strategy profile $z^j$ differs from the strategy profile $z^{j-1}$ for each $j \in \{1, 2, \cdots, |\Hat{\mcL}(i)|\}$ only for player $l_j \in \Hat{\mcL}(i)$. Hence:
\begin{align*}
    \Psi(\Tilde{z})-\Psi(\Hat{z})&=\sum_{j=1}^{|\Hat{\mcL}(i)|} \big( \Psi(z^j)-\Psi(z^{j-1}) \big) \\
    &= \sum_{j=1}^{|\Hat{\mcL}(i)|}  \big( Y_{l_j}\big(z_{l_j}^j, z_{\mcN_{l_j}^p}^j\big) -Y_{l_j}\big(z_{l_j}^{j-1}, z_{\mcN_{l_j}^p}^{j-1}\big) \big) \\
    &=\sum_{l \in \Hat{\mcL}(i)} \big( Y_l\big(\Tilde{z}_l, \Tilde{z}_{\mcN_l^{\tp}}\big)-Y_l\big(\Hat{z}_l, \Hat{z}_{\mcN_l^{\tp}}\big) \big) \\
    &=\sum_{l \in \Hat{\mcL}(i)} Y_l\big(\Tilde{z}_l, \Tilde{z}_{\mcN_l^{\tp}}\big) -\sum_{l \in \Hat{\mcL}(i)} Y_l\big(\Hat{z}_l, \Hat{z}_{\mcN_l^{\tp}}\big)
\end{align*}   
\end{IEEEproof}

\begin{IEEEproof}[Proof of Theorem \ref{Theorem: LIG attain global opt. of PSI}]
    Recall from Lemma \ref{Lemma: stationary distribution LIG} that $\pi (z)=\frac{e^{\beta \Psi (z)}}{\sum_{z' \in \mcZ_{\mcL}} e^{\beta \Psi (z')}}, \, \forall z \in \mcZ_{\mcL}$ is the unique stationary distribution achieved by Algorithm \ref{Algorithm: LIG based UE and Beam Selection} over the strategy space.
    Recall that  $\mcZ_{\mcL}^{\topt}$ denotes a set of strategies of the players, which achieves the global optimum of $\Psi$ over the strategy space.
    For sufficiently large $\beta$, $e^{\beta \Psi(z^{\topt})} \gg e^{\beta \Psi(z)}, \, \forall z^{\topt} \in \mcZ_{\mcL}^{\topt},  z \in \mcZ_{\mcL} \setminus \mcZ_{\mcL}^{\topt}$.
     Hence:
    \begin{align*}
    \lim_{\beta \uparrow \infty} \pi(z)=
        \begin{cases}
           \frac{1}{|\mcZ_{\mcL}^{\topt}|},& \forall z  \in \mcZ_{\mcL}^{\topt}, \\
           0, & z \in \mcZ_{\mcL} \setminus \mcZ_{\mcL}^{\topt}.
        \end{cases}
    \end{align*}
    Hence, Algorithm \ref{Algorithm: LIG based UE and Beam Selection} achieves the global optimum of 
    $\Psi$.
\end{IEEEproof}
\section*{Proof of Theorem \ref{Theorem: LIG Algo achieve optimal UE and beam selection}}
Let $\mcM_l^{\tc}:=\{ i' \in \mcA|{l'}_{\ta}=i', z_{l'}>0, l' \in \mcN_{l}^{\tc} \}$ denote the set of APs, at least one of whose active players, can get interference from player $l$.  Let $\mcM_i^{\tac}:=\bigcup_{l \in \mcP_i} \mcM_l^{\tc}$ denote the set of APs, at least one of whose active players, can get interference from at least one of the active players at AP $i$.  Let $\mcM_i^{\ag}:=\bigcup_{l \in \mcP_i} \mcM_l^{\tg}$ denote the set of APs, at least one of whose active players, can create interference to at least one of the active players at AP $i$. Recall that $\mcM_l^{\tg}:=\{ i' \in \mcA|{l'}_{\ta}=i', z_{l'}>0, l' \in \mcN_{l}^{\tg}\}$ denotes the set of APs, at least one of whose active players, can create interference to the player $l$.

\begin{IEEEproof}[Proof of Lemma \ref{Lemma: Strategy Satisfy Constraints}]
If $\Hat{z} \in \{ \mcZ_{\mcL} \setminus \mcZ_{\mcL}^{\tc} \}$, then at least one of the two conditions: (i) $ |\{ l \in \mcL| l_{\ta}=i^*, \Hat{z}_l>0 \}| > 1$ for some  $i^* \in \mcA$, and (ii) $|\{ l \in \mcL| l_{\tu}=u^*, \Hat{z}_l>0 \}| > 1$  for some  $u^* \in \mcU$ must hold. For both cases, we show that $\Psi(\Tilde{z})>\Psi(\Hat{z})$.

Lets consider the first case where more than one player are on at some AP. Suppose $ |\{ l \in \mcL| l_{\ta}=i^*, \Hat{z}_l>0 \}| > 1$ for  $i^* \in \mcA$.   

\begin{align*}
    \Psi(\Hat{z})=&\sum_{l \in \mcL} V_l(\Hat{z}_l,\Hat{z}_{\mcN_l^{\tut}})\\
    =&\sum_{l \in \mcL \setminus \bigcup_{i \in \{ i^* \bigcup \mcM_{i^*}^{\tac} \} } \mcP_i } V_l(\Hat{z}_l,\Hat{z}_{\mcN_l^{\tut}})\\
    &+\sum_{l \in \bigcup_{i \in \{ i^* \bigcup \mcM_{i^*}^{\tac} \} } \mcP_i  } V_l(\Hat{z}_l,\Hat{z}_{\mcN_l^{\tut}})
\end{align*}
Note that the first term in the second equality does not get affected by the actions of the players at AP $i^*$. Denote the second term in the equation as $\widehat{\text{ST}}$ and note:
\begin{align*}
\widehat{\text{ST}}=& \frac{1}{k_{i^*}} \sum_{l \in \mcP_{i^*}} \bigg\{ \bigg[ \frac{1}{\prod_{i' \in \mcM_{i^*}^{\ag} } k_{i'}} \sum_{ \mcP \in \times_{i' \in \mcM_{i^*}^{\ag} } \mcP_{i'} } R(l,\mcP) \bigg] \\
&+ \bigg[ \sum_{i' \in \mcM_{i^*}^{\tac} } \frac{1}{\prod_{i'' \in \mcM_{i'}^{\ag} \setminus i^* } k_{i''}} \\
&\sum_{l' \in \mcP_{i'}} \sum_{ \mcP \in \times_{i'' \in \mcM_{i'}^{\ag} \setminus i^* } \mcP_{i''} } R(l',l,\mcP)  \bigg] \bigg\} 
\end{align*}
Let $\widehat{\text{ST}}(l), l \in \mcP_{i^*}$, denote the term inside the curly brackets in the above equation. Note that this term does not depend on the actions of players $l' \in \mcP_{i^*} \setminus l$. So if we switch off a player $l^* \in \mcP_{i^*}$, then the second term becomes $\widetilde{\text{ST}}=\frac{1}{k_{i^*}-1} \sum_{l \in \mcP_{i^*} \setminus l^* } \widehat{\text{ST}}(l)$. Also, $\widehat{\text{ST}}$ is the average of $\widehat{\text{ST}}(l)$ over all $l \in \mcP_{i^*}$, whereas  $\widetilde{\text{ST}}$ is the average of $\widehat{\text{ST}}(l)$ over $l \in \mcP_{i^*} \setminus l^*$. If $l^*=\argmin_{l \in \mcP_{i^*}} \widehat{\text{ST}}(l)$, then $\widetilde{\text{ST}}>\widehat{\text{ST}}$; this implies $\Psi(\Tilde{z})>\Psi(\Hat{z})$ with $\Tilde{z}_l=\Hat{z}_l, \forall l \in \mcL\setminus l^*$ and $\Tilde{z}_{l^*}=0$.

Now, let us consider the second case where more than one player are on at some UE. Suppose $|\{ l \in \mcL| l_{\tu}=u^*, \Hat{z}_l>0 \}| > 1$ for  $u^* \in \mcU$. Note that when more than one player corresponds to the same UE, then the utility of each of the players corresponding to the same UE becomes zero. Hence, when one of them gets switched off, either the sum of the utilities of the players gets improved or remains unchanged. When $u^*$ has exactly two players, then switching off any of these two improves the sum of the utilities of the players. Suppose $l^*$ gets switched off. Then, $\Psi(\Tilde{z}) \ge \Psi(\Hat{z})$ with $\Tilde{z}_l=\Hat{z}_l, \forall l \in \mcL\setminus l^*$ and $\Tilde{z}_{l^*}=0$.
\end{IEEEproof}
Note that $\mcF$ (respectively, $\Breve{\mcA}$) in (\ref{Eq: Objective Main}) is $\{ <l_{\tb},l_{\tu},l_{\ta}>| l \in \mcL, z_l>0 \}$ (respectively, $\{ l_{\ta}| l \in \mcL, z_l>0 \}$)  given $z \in \mcZ_{\mcL}^{\tc}$.
\begin{lemma} \label{Lemma: Psi equal sum weighted rate}
If $z \in \mcZ_{\mcL}^{\tc}$, then $\Psi(z)=R_{\mcS,\bw,\mcF}$.
\end{lemma}
\begin{IEEEproof}[Proof of Lemma \ref{Lemma: Psi equal sum weighted rate}]
We can write the sum of the weighted rates of the UEs, $R_{\mcS,\bw,\mcF}$, in terms of $z$ as follows.
\begin{align*}
&\sum_{l \in \{l'|z_{l'}>0 \}} w_{l_{\tu}} B \log_2 \Bigg( 1+ \frac{S_{l_{\tb},l_{\tu},l_{\ta}}}{N_0+\sum_{l'\in \{l''|z_{l''}>0 \} \setminus l }  S_{{l'}_b,l_{\tu},{l'}_{\ta}} } \Bigg)\\
=& \sum_{l \in \{l'|z_{l'}>0 \}} w_{l_{\tu}} B \log_2 \Bigg( 1+ \frac{S_{l_{\tb},l_{\tu},l_{\ta}}}{N_0+\sum_{l'\in \mcN_l^{\tg}}  S_{{l'}_b,l_{\tu},{l'}_{\ta}} } \Bigg).
\end{align*}
The above equality holds since player $l$ only gets interference from players $l'\in \mcN_l^{\tg}$. It is easy to see that $k_i=1, \forall l \in \{l' \in \mcL| z_{l'}>0 \}$. Also, $\mcN_l^{\tut}=\mcN_l^{\tg}$. So, $\prod_{i \in \{ l_{\ta} \bigcup \mcM_{l_{\ta}}^{\ag} \} } k_i=1$ and $\times_{i \in \mcM_{l_{\ta}}^{\ag}} \mcP_i=\mcN_l^{\tg}$. Hence, $\Psi(z)$
\begin{align*}
    &=\sum_{l \in \mcL} V_l(z_l, z_{\mcN_l^{\tut}})\\
    &=\sum_{l \in \{ l' \in \mcL|z_{l'}>0 \}} w_{l_{\tu}} B \log_2 \Bigg( 1+ \frac{S_{l_{\tb},l_{\tu},l_{\ta}}}{N_0+\sum_{l'\in \mcN_l^{\tg}}  S_{{l'}_b,l_{\tu},{l'}_{\ta}} } \Bigg). 
\end{align*}
\end{IEEEproof}
Lemma \ref{Lemma: Psi equal sum weighted rate} states that the sum of the utilities of the players will be the same as the sum of the weighted rates of the UEs corresponding to the active players when the strategies of the players satisfy the constraints of Problem \ref{Problem: Original UE and beam selection Problem}.
\begin{IEEEproof}[Proof of Theorem \ref{Theorem: LIG Algo achieve optimal UE and beam selection}]
Note that Lemma \ref{Lemma: Strategy Satisfy Constraints} implies that if more than one player remains active at any AP or UE, then there exists a player corresponding to an AP or a UE having two or more incident active players such that switching that player off leads to a strategy profile of players whose value of $\Psi$ is either the same as before or higher. Hence, by Theorem \ref{Theorem: LIG attain global opt. of PSI}, for sufficiently large $\beta$, the strategy profile of the players selected by  Algorithm \ref{Algorithm: LIG based UE and Beam Selection} must belong to $\mcZ_{\mcL}^{\tc}$. Note that a strategy profile of players $z \in \mcZ_{\mcL}^{\tc}$ satisfies the constraints specified in Problem \ref{Problem: Original UE and beam selection Problem}. If $z \in \mcZ_{\mcL}^{\tc}$, then Lemma \ref{Lemma: Psi equal sum weighted rate} states that the sum of the utilities of the players, $\Psi(z)$, is the same as the sum of the weighted rates of the UEs corresponding to the players. Therefore, Algorithm \ref{Algorithm: LIG based UE and Beam Selection} achieves the optimal  UE and beam selection of Problem \ref{Problem: Original UE and beam selection Problem}.
\end{IEEEproof}
\section*{Proof of Theorem \ref{Theorem: No. of iter. to reach optimum}}
\begin{IEEEproof}[Proof of Theorem \ref{Theorem: No. of iter. to reach optimum}]
     By definition of $D$, we have that for $i>D$: $\nu_i$
    \begin{align*}
        =&\Pr(\widetilde{Z}(i) \in \mcZ_{\mcL}^{\topt}|\widetilde{Z}(i-D) \in \mcZ_{\mcL}^{\topt}) \Pr(\widetilde{Z}(i-D) \in \mcZ_{\mcL}^{\topt}) \\
        &+ \Pr(\widetilde{Z}(i) \in \mcZ_{\mcL}^{\topt}|\widetilde{Z}(i-D) \in \mcZ_{\mcL} \setminus \mcZ_{\mcL}^{\topt} ) \times \\
        &\Pr(\widetilde{Z}(i-D) \in \mcZ_{\mcL} \setminus \mcZ_{\mcL}^{\topt}) \\
        =& \nu_{i-D}+(1-\nu_{i-D}) \Pr(\widetilde{Z}(i) \in \mcZ_{\mcL}^{\topt}|\widetilde{Z}(i-D) \in \mcZ_{\mcL} \setminus \mcZ_{\mcL}^{\topt} )\\
        \ge & \nu_{i-D}+(1-\nu_{i-D}) \eta^D.
    \end{align*}
    Thus, $(1-\nu_i) \le (1-\nu_{i-D})(1-\eta^D)$. Let $\tau$ denote the iteration in which the chain goes to an absorbing state for the first time. Then $\tau=i$ will be the event $\{ \widetilde{Z}(i) \in \mcZ_{\mcL}^{\topt} \} \bigcap_{i'=0}^{i-1} \{ \widetilde{Z}(i') \in \mcZ_{\mcL} \setminus \mcZ_{\mcL}^{\topt} \}$.
     Note that for the absorbing Markov chain, we can write $\Pr(\tau >i)=1-\nu_i$. 
     Let $r=\lfloor \frac{i}{D}\rfloor$. 
     Then, $1-v_i \le (1-\eta^D)^r (1-\nu_{i-Dr})\le (1-\eta^D)^r (1-\nu_0)$. 
    \begin{align*}
        \mathbf{E}[\tau]&=\sum_{i=0}^{\infty} \Pr(\tau > i)=\sum_{i=0}^{\infty} (1-\nu_i)\\
        &\le \sum_{r=0}^{\infty} \sum_{d=0}^{D-1} (1-\eta^D)^r (1-\nu_0)\\
        &=\frac{D(1-\nu_0)}{\eta^D}.
    \end{align*}    
    The result follows.
\end{IEEEproof}
\bibliography{reference} 

\begin{thebibliography}{10}
\providecommand{\url}[1]{#1}
\csname url@samestyle\endcsname
\providecommand{\newblock}{\relax}
\providecommand{\bibinfo}[2]{#2}
\providecommand{\BIBentrySTDinterwordspacing}{\spaceskip=0pt\relax}
\providecommand{\BIBentryALTinterwordstretchfactor}{4}
\providecommand{\BIBentryALTinterwordspacing}{\spaceskip=\fontdimen2\font plus
\BIBentryALTinterwordstretchfactor\fontdimen3\font minus \fontdimen4\font\relax}
\providecommand{\BIBforeignlanguage}[2]{{%
\expandafter\ifx\csname l@#1\endcsname\relax
\typeout{** WARNING: IEEEtran.bst: No hyphenation pattern has been}%
\typeout{** loaded for the language `#1'. Using the pattern for}%
\typeout{** the default language instead.}%
\else
\language=\csname l@#1\endcsname
\fi
#2}}
\providecommand{\BIBdecl}{\relax}
\BIBdecl

\bibitem{singh2023ngub}
S.~K. Singh, S.~Sahu, A.~Thawait, P.~Chaporkar, and G.~S. Kasbekar, ``{NGUB}: {Novel} greedy algorithms for user and beam selection in {mmWave} networks,'' in \emph{2023 Fourteenth International Conference on Ubiquitous and Future Networks (ICUFN)}.\hskip 1em plus 0.5em minus 0.4em\relax IEEE, 2023, pp. 421--426.

\bibitem{cisco2020}
\BIBentryALTinterwordspacing
Cisco, ``{Cisco Annual Internet Report (2018–2023) White Paper}.'' [Online]. Available: \url{https://www.cisco.com/c/en/us/solutions/collateral/executive-perspectives/annual-internet-report/white-paper-c11-741490.html}
\BIBentrySTDinterwordspacing

\bibitem{andrews2014what}
J.~G. Andrews, S.~Buzzi, W.~Choi, S.~V. Hanly, A.~Lozano, A.~C. Soong, and J.~C. Zhang, ``What will {5G} be?'' \emph{IEEE Journal on Selected Areas in Communications}, vol.~32, no.~6, pp. 1065--1082, 2014.

\bibitem{agiwal2016next}
M.~Agiwal, A.~Roy, and N.~Saxena, ``Next generation {5G} wireless networks: {A} comprehensive survey,'' \emph{IEEE Communications Surveys \& Tutorials}, vol.~18, no.~3, pp. 1617--1655, 2016.

\bibitem{niu2015survey}
Y.~Niu, Y.~Li, D.~Jin, L.~Su, and A.~V. Vasilakos, ``A survey of millimeter wave {(mmWave)} communications for {5G}: {Opportunities} and challenges,'' \emph{Wireless Networks}, vol.~21, no.~8, pp. 2657--2676, 2015.

\bibitem{sakaguchi2017where}
K.~Sakaguchi, T.~Haustein, S.~Barbarossa, E.~C. Strinati, A.~Clemente, G.~Destino, A.~P{\"a}rssinen, I.~Kim, H.~Chung, J.~Kim \emph{et~al.}, ``Where, when, and how {mmWave} is used in {5G} and beyond,'' \emph{IEICE Transactions on Electronics}, vol. 100, no.~10, pp. 790--808, 2017.

\bibitem{liu2016user}
D.~Liu, L.~Wang, Y.~Chen, M.~Elkashlan, K.-K. Wong, R.~Schober, and L.~Hanzo, ``User association in {5G} networks: {A} survey and an outlook,'' \emph{IEEE Communications Surveys \& Tutorials}, vol.~18, no.~2, pp. 1018--1044, 2016.

\bibitem{attiah2020survey}
M.~L. Attiah, A.~A.~M. Isa, Z.~Zakaria, M.~Abdulhameed, M.~K. Mohsen, and I.~Ali, ``A survey of {mmWave} user association mechanisms and spectrum sharing approaches: an overview, open issues and challenges, future research trends,'' \emph{Wireless Networks}, vol.~26, no.~4, pp. 2487--2514, 2020.

\bibitem{sur2016beamspy}
S.~Sur, X.~Zhang, P.~Ramanathan, and R.~Chandra, ``{BeamSpy}: Enabling robust 60 {GHz} links under blockage,'' in \emph{13th USENIX Symposium on Networked Systems Design and Implementation (NSDI 16)}, 2016, pp. 193--206.

\bibitem{zhou2017beamforecast}
A.~Zhou, X.~Zhang, and H.~Ma, ``Beam-forecast: {Facilitating} mobile 60 {GHz} networks via model-driven beam steering,'' in \emph{IEEE INFOCOM 2017-IEEE Conference on Computer Communications}.\hskip 1em plus 0.5em minus 0.4em\relax IEEE, 2017, pp. 1--9.

\bibitem{haider2018listeer}
M.~K. Haider, Y.~Ghasempour, D.~Koutsonikolas, and E.~W. Knightly, ``Listeer: {Mmwave} beam acquisition and steering by tracking indicator {LEDs} on wireless {APs},'' in \emph{Proceedings of the 24th Annual International Conference on Mobile Computing and Networking}, 2018, pp. 273--288.

\bibitem{wei2017pose}
T.~Wei and X.~Zhang, ``Pose information assisted 60 {GHz} networks: Towards seamless coverage and mobility support,'' in \emph{Proceedings of the 23rd Annual International Conference on Mobile Computing and Networking}, 2017, pp. 42--55.

\bibitem{jog2019many}
S.~Jog, J.~Wang, J.~Guan, T.~Moon, H.~Hassanieh, and R.~R. Choudhury, ``{Many-to-Many} beam alignment in millimeter wave networks,'' in \emph{16th USENIX Symposium on Networked Systems Design and Implementation (NSDI 19)}, 2019, pp. 783--800.

\bibitem{yang2020mmmuxing}
Y.~Yang, A.~Zhou, D.~Xu, S.~Yang, L.~Wu, H.~Ma, T.~Wei, and J.~Liu, ``{mmMuxing}: {Pushing} the limit of spatial reuse in directional millimeter-wave wireless networks,'' in \emph{2020 17th Annual IEEE International Conference on Sensing, Communication, and Networking (SECON)}.\hskip 1em plus 0.5em minus 0.4em\relax IEEE, 2020, pp. 1--9.

\bibitem{yang2021mdsr}
Y.~Yang, A.~Zhou, D.~Xu, K.~Liang, H.~Ma, T.~Wei, and J.~Liu, ``{MDSR}: {Multi}-dimensional spatial reuse enhancement for directional millimeter-wave wireless networks,'' \emph{IEEE Transactions on Mobile Computing}, vol.~21, no.~12, pp. 4439--4455, 2022.

\bibitem{zhang2022reinforcement}
Y.~Zhang and R.~W. Heath, ``Reinforcement learning-based joint user scheduling and link configuration in millimeter-wave networks,'' \emph{IEEE Transactions on Wireless Communications}, pp. 1--17, 2022.

\bibitem{sha2023versatile}
Z.~Sha, Y.~Ming, C.~Sun, and Z.~Wang, ``Versatile resource management for millimeter-wave cellular network: {Near} interference-free scheduling methodology,'' \emph{IEEE Transactions on Vehicular Technology}, pp. 1--16, 2023.

\bibitem{kleinberg2006algorithm}
J.~Kleinberg and E.~Tardos, \emph{Algorithm design}.\hskip 1em plus 0.5em minus 0.4em\relax Pearson Education India, 2006.

\bibitem{kim2011distributed}
H.~Kim, G.~De~Veciana, X.~Yang, and M.~Venkatachalam, ``Distributed {$\alpha$}-optimal user association and cell load balancing in wireless networks,'' \emph{IEEE/ACM Transactions on Networking}, vol.~20, no.~1, pp. 177--190, 2011.

\bibitem{madan2010cell}
R.~Madan, J.~Borran, A.~Sampath, N.~Bhushan, A.~Khandekar, and T.~Ji, ``Cell association and interference coordination in heterogeneous {LTE-- A} cellular networks,'' \emph{IEEE Journal on Selected Areas in Communications}, vol.~28, no.~9, pp. 1479--1489, 2010.

\bibitem{awais2017efficient}
M.~Awais, A.~Ahmed, M.~Naeem, M.~Iqbal, W.~Ejaz, A.~Anpalagan, and H.~S. Kim, ``Efficient joint user association and resource allocation for cloud radio access networks,'' \emph{IEEE Access}, vol.~5, pp. 1439--1448, 2017.

\bibitem{ye2013user}
Q.~Ye, B.~Rong, Y.~Chen, M.~Al-Shalash, C.~Caramanis, and J.~G. Andrews, ``User association for load balancing in heterogeneous cellular networks,'' \emph{IEEE Transactions on Wireless Communications}, vol.~12, no.~6, pp. 2706--2716, 2013.

\bibitem{mesodiakaki2016energy}
A.~Mesodiakaki, F.~Adelantado, L.~Alonso, M.~Di~Renzo, and C.~Verikoukis, ``Energy-and spectrum-efficient user association in millimeter-wave backhaul small-cell networks,'' \emph{IEEE Transactions on Vehicular Technology}, vol.~66, no.~2, pp. 1810--1821, 2016.

\bibitem{khawam2020coordinated}
K.~Khawam, S.~Lahoud, M.~El~Helou, S.~Martin, and F.~Gang, ``Coordinated framework for spectrum allocation and user association in {5G HetNets with mmWave},'' \emph{IEEE Transactions on Mobile Computing}, vol.~21, no.~4, pp. 1226--1243, 2020.

\bibitem{skouroumounis2017low}
C.~Skouroumounis, C.~Psomas, and I.~Krikidis, ``Low-complexity base station selection scheme in {mmWave} cellular networks,'' \emph{IEEE Transactions on Communications}, vol.~65, no.~9, pp. 4049--4064, 2017.

\bibitem{zhang2017energy}
H.~Zhang, S.~Huang, C.~Jiang, K.~Long, V.~C. Leung, and H.~V. Poor, ``Energy efficient user association and power allocation in millimeter-wave-based ultra dense networks with energy harvesting base stations,'' \emph{IEEE Journal on Selected Areas in Communications}, vol.~35, no.~9, pp. 1936--1947, 2017.

\bibitem{athanasiou2013auction}
G.~Athanasiou, P.~C. Weeraddana, and C.~Fischione, ``Auction-based resource allocation in millimeter-wave wireless access networks,'' \emph{IEEE Communications Letters}, vol.~17, no.~11, pp. 2108--2111, 2013.

\bibitem{liu2019joint}
R.~Liu, Q.~Chen, G.~Yu, and G.~Y. Li, ``Joint user association and resource allocation for multi-band millimeter-wave heterogeneous networks,'' \emph{IEEE Transactions on Communications}, vol.~67, no.~12, pp. 8502--8516, 2019.

\bibitem{soleimani2018cluster}
B.~Soleimani and M.~Sabbaghian, ``Cluster-based resource allocation and user association in {mmWave} femtocell networks,'' \emph{IEEE Transactions on Communications}, vol.~68, no.~3, pp. 1746--1759, 2018.

\bibitem{xu2019ping}
H.~Xu, J.~Yu, and S.~Zhu, ``Ping-pong optimization of user selection and beam allocation for millimeter wave communications,'' \emph{IEEE Access}, vol.~7, pp. 133\,178--133\,189, 2019.

\bibitem{hegde2019matching}
A.~Hegde and K.~Srinivas, ``Matching theoretic beam selection in millimeter-wave multi-user {MIMO} systems,'' \emph{IEEE Access}, vol.~7, pp. 25\,163--25\,170, 2019.

\bibitem{cheng2020low}
Z.~Cheng, Z.~Wei, and H.~Yang, ``Low-complexity joint user and beam selection for beamspace {mmWave MIMO} systems,'' \emph{IEEE Communications Letters}, vol.~24, no.~9, pp. 2065--2069, 2020.

\bibitem{ahn2022machine}
H.~Ahn, I.~Orikumhi, J.~Kang, H.~Park, H.~Jwa, J.~Na, and S.~Kim, ``Machine learning-based vision-aided beam selection for {mmWave} multiuser {MISO} system,'' \emph{IEEE Wireless Communications Letters}, vol.~11, no.~6, pp. 1263--1267, 2022.

\bibitem{alizadeh2022reinforcement}
A.~Alizadeh and M.~Vu, ``Reinforcement learning for user association and handover in {mmWave}-enabled networks,'' \emph{IEEE Transactions on Wireless Communications}, vol.~21, no.~11, pp. 9712--9728, 2022.

\bibitem{zhang2021non}
X.~Zhang, S.~Sarkar, A.~Bhuyan, S.~K. Kasera, and M.~Ji, ``A non-cooperative game-based distributed beam scheduling framework for {5G} millimeter-wave cellular networks,'' \emph{IEEE Transactions on Wireless Communications}, vol.~21, no.~1, pp. 489--504, 2021.

\bibitem{wang2023joint}
L.~Wang, B.~Ai, Y.~Niu, H.~Jiang, S.~Mao, Z.~Zhong, and N.~Wang, ``Joint user association and transmission scheduling in integrated {mmWave} access and terahertz backhaul networks,'' \emph{IEEE Transactions on Vehicular Technology}, vol.~72, no.~12, pp. 15\,930--15\,940, 2023.

\bibitem{samimi20163d}
M.~K. Samimi and T.~S. Rappaport, ``{3-D} millimeter-wave statistical channel model for {5G} wireless system design,'' \emph{IEEE Transactions on Microwave Theory and Techniques}, vol.~64, no.~7, pp. 2207--2225, 2016.

\bibitem{fleischner2010maximum}
H.~Fleischner, G.~Sabidussi, and V.~I. Sarvanov, ``Maximum independent sets in 3-and 4-regular {Hamiltonian} graphs,'' \emph{Discrete Mathematics}, vol. 310, no.~20, pp. 2742--2749, 2010.

\bibitem{west2001introduction}
D.~B. West \emph{et~al.}, \emph{Introduction to graph theory}.\hskip 1em plus 0.5em minus 0.4em\relax Prentice hall Upper Saddle River, 2001, vol.~2.

\bibitem{swenson2018best}
B.~Swenson, R.~Murray, and S.~Kar, ``On best-response dynamics in potential games,'' \emph{SIAM Journal on Control and Optimization}, vol.~56, no.~4, pp. 2734--2767, 2018.

\bibitem{xu2011opportunistic}
Y.~Xu, J.~Wang, Q.~Wu, A.~Anpalagan, and Y.-D. Yao, ``Opportunistic spectrum access in cognitive radio networks: {Global} optimization using local interaction games,'' \emph{IEEE Journal of Selected Topics in Signal Processing}, vol.~6, no.~2, pp. 180--194, 2011.

\bibitem{frigessi1993convergence}
A.~Frigessi, P.~Stefano, C.-R. Hwang, and S.-J. Sheu, ``Convergence rates of the gibbs sampler, the metropolis algorithm and other single-site updating dynamics,'' \emph{Journal of the Royal Statistical Society Series B: Statistical Methodology}, vol.~55, no.~1, pp. 205--219, 1993.

\bibitem{roberts1994simple}
G.~O. Roberts and A.~F. Smith, ``Simple conditions for the convergence of the {Gibbs} sampler and {Metropolis-Hastings} algorithms,'' \emph{Stochastic Processes and their Applications}, vol.~49, no.~2, pp. 207--216, 1994.

\bibitem{young1998individual}
H.~P. Young, \emph{Individual strategy and social structure: {An} evolutionary theory of institutions}.\hskip 1em plus 0.5em minus 0.4em\relax Princeton University Press, 1998.

\bibitem{marden2009cooperative}
J.~R. Marden, G.~Arslan, and J.~S. Shamma, ``Cooperative control and potential games,'' \emph{IEEE Transactions on Systems, Man, and Cybernetics, Part B (Cybernetics)}, vol.~39, no.~6, pp. 1393--1407, 2009.

\bibitem{zhang2011weighted}
H.~Zhang, L.~Venturino, N.~Prasad, P.~Li, S.~Rangarajan, and X.~Wang, ``Weighted sum-rate maximization in multi-cell networks via coordinated scheduling and discrete power control,'' \emph{IEEE Journal on Selected Areas in Communications}, vol.~29, no.~6, pp. 1214--1224, 2011.

\bibitem{zheng2015optimal}
J.~Zheng, Y.~Cai, X.~Chen, R.~Li, and H.~Zhang, ``Optimal base station sleeping in green cellular networks: {A} distributed cooperative framework based on game theory,'' \emph{IEEE Transactions on Wireless Communications}, vol.~14, no.~8, pp. 4391--4406, 2015.

\bibitem{liu2018decentralized}
Y.~Liu, X.~Fang, M.~Xiao, and S.~Mumtaz, ``Decentralized beam pair selection in multi-beam millimeter-wave networks,'' \emph{IEEE Transactions on Communications}, vol.~66, no.~6, pp. 2722--2737, 2018.

\bibitem{ermon2014designing}
S.~Ermon, C.~Gomes, A.~Sabharwal, and B.~Selman, ``Designing fast absorbing {Markov} chains,'' in \emph{Proceedings of the AAAI Conference on Artificial Intelligence}, vol.~28, no.~1, 2014.

\bibitem{ju20203d}
S.~Ju, Y.~Xing, O.~Kanhere, and T.~S. Rappaport, ``{3-D} statistical indoor channel model for millimeter-wave and sub-terahertz bands,'' in \emph{GLOBECOM 2020-2020 IEEE Global Communications Conference}.\hskip 1em plus 0.5em minus 0.4em\relax IEEE, 2020, pp. 1--7.

\bibitem{ju2021millimeter}
------, ``Millimeter wave and sub-terahertz spatial statistical channel model for an indoor office building,'' \emph{IEEE Journal on Selected Areas in Communications}, vol.~39, no.~6, pp. 1561--1575, 2021.

\bibitem{jain1984quantitative}
R.~K. Jain, D.-M.~W. Chiu, W.~R. Hawe \emph{et~al.}, ``A quantitative measure of fairness and discrimination,'' \emph{Eastern Research Laboratory, Digital Equipment Corporation, Hudson, MA}, vol.~21, 1984.

\end{thebibliography}
\bibliographystyle{IEEEtran}
\end{document}